\documentclass{nature}
\usepackage{graphicx}
\usepackage{dcolumn}
\usepackage{bm}
\usepackage{amsmath}
\usepackage{amssymb}
\usepackage{color}
\usepackage{tikz}
\usepackage{caption}
\usepackage{lineno}
\title{Observation of a symmetry-protected topological phase in external magnetic fields} 

\author{Zhihuang Luo${^{1}}^{*}$, Wenzhao Zhang$^{2}$, Xinfang Nie$^{3}$, \& Dawei Lu${^{3}}^{*}$}

\begin{document}

\maketitle

\begin{affiliations}

 \item Guangdong Provincial Key Laboratory of Quantum Metrology and Sensing \& School of Physics and Astronomy, Sun Yat-Sen University (Zhuhai Campus), Zhuhai 519082, China.
 \item Department of Physics, Ningbo University, Ningbo 315211, China.
 \item Shenzhen Institute for Quantum Science and Engineering and Department of Physics, Southern University of Science and Technology, Shenzhen 518055, China.

\end{affiliations}

\begin{abstract}
Topological phases have greatly improved our understanding of modern conception of phases of matter that go beyond the paradigm of symmetry breaking and are not described by local order parameters. Instead, characterization of topological phases requires the robust topological invariants, whereas measuring these global quantities presents an outstanding challenge for experiments, especially in interacting systems. Here we report the real-space observation of a symmetry-protected topological phase with interacting nuclear spins, and probe the interaction-induced transition between two topologically distinct phases, both of which are classified by many-body Chern numbers obtained from the dynamical response to the external magnetic fields. The resulting value of Chern number ($\bar{\mathcal{C}h} = 1.958 \pm 0.070$) demonstrates the robust ground state degeneracy, and also determines the number of nontrivial edge states. Our findings enable direct characterization of topological features of quantum many-body states through gradually decreasing the strength of the introduced external fields.
\end{abstract}

Symmetry plays a fundamental role in classifying different phases of matter and their associated phase transitions, where the symmetry is spontaneously broken \cite{Landau1937TheoryI}. However, the discovery of topological phases goes beyond this symmetry breaking paradigm, whose characterization requires the topological invariants, such as the Chern numbers, that are robust against local perturbations \cite{Wen1990Topological,Luo2018Experimentally}. A prominent example is the integer quantum Hall effect \cite{Klitzing1980New}, demonstrating that the Hall conductance in response to an external electric field is quantized by integer Chern numbers. Recent progress of topological insulators and superconductors \cite{Hasan2010Colloquium,Qi2011Topological} has led to the broad notion of symmetry-protected topological (SPT) phases \cite{Gu2009Tensor,Chen2012Symmetry}, which are short-range entangled quantum states with the gapless or degenerate edge modes protected by a symmetry. The edge states of SPT phases have been observed in a variety of physical platforms \cite{Wang2009Observation,Rechtsman2013Photonic,Susstrunk2015Observation,Hafezi2013Imaging,Meier2016Observation,Leder2016Real-space,Zhang2019Experimental,de2019Observation}, and the momentum-resolved Berry curvature and further relevant Chern number, as the central measure of topology, have also been used to determine the band structures of noninteracting SPT phases at the single-particle level \cite{Flaschner2016Experimental,Aidelsburger2015Measuring,Li2016Bloch,Atala2013Direct,Wang2019Direct}. Whereas in quantum many-body systems with interactions, where there may exist a large number of SPT phases \cite{Chen2012Symmetry}, the underlying Chern number is not easily accessible. Because the interactions complicate the energy levels in real space, and make it difficult to fully determine the band eigenstates based on the previous methods of momentum space \cite{Flaschner2016Experimental,Aidelsburger2015Measuring,Li2016Bloch,Atala2013Direct,Wang2019Direct}. 

Going beyond the noninteracting systems, we report the real-space measurement of many-body Chern numbers, and directly observe an interacting SPT phase. The experiment is performed on the sample of trans-crotonic acid, which consists of a staggered one-dimensional (1D) spin-$1/2$ chain of interacting carbon nuclei. We use this system to realize a 1D Su-Schrieffer-Heeger (SSH) model with an SPT phase \cite{Su1979Solitons,Asboth2016A,Choo2018Measurement}. By introducing the external magnetic fields to the SSH model, the Berry curvatures are allow to be defined in the parameter space of these fields, and can be extracted from the linear response of spin magnetizations to the quench velocity \cite{Gritsev2012Dynamical,Roushan2014Observation,Schroer2014Measuring,Luo2016Experimental}. The relevant Chern number obtained by the integral of Berry curvature determines the number of degenerate ground states, and also the number of nontrivial edge states according to the bulk-boundary correspondence. We characterize the two topologically distinct phases, with the features of nonzero Chern number for the topological phase and zero value for the trivial one, as the introduced external magnetic field decreases to zero. Furthermore, we probe the interaction-induced transition from a topological phase to a trivial one, under a fixed small external field. Our measurements demonstrate clear signatures of the SPT phase, and establish a general approach for exploring the topological features of interacting quantum many-body systems.

The SSH model is a paradigmatic example with nontrivial topology, originally describing electrons hopping on a 1D chain with staggered hopping amplitudes. We consider the spin Hamiltonian of 1D SSH model derived via a Jordan-Wigner transformation \cite{SM2}
\begin{equation}\label{eq: spinHamofSSH}
    \mathcal{H}_{\text{SSH}}^{\text{S}}= -J_1\sum_{i=1}^N (\sigma_{2i-1}^x\sigma_{2i}^x +\sigma_{2i-1}^y\sigma_{2i}^y) - J_2 \sum_{i=1}^{N-1} (\sigma_{2i}^x\sigma_{2i+1}^x +\sigma_{2i}^y\sigma_{2i+1}^y),
\end{equation}
which consists of $N$ unit cells with each unit cell hosting two spins. The $\sigma_i^x$ $(\sigma_i^y)$ stand for Pauli matrices. Depending on the ratio of $J_1$ and $J_2$, the chain exhibits two topologically distinct phases: a topological one with four-fold degenerate ground states if $J_1 < J_2$, and a trivial one with a single ground state if $J_1 > J_2$. 

\begin{figure}
\centering
\includegraphics[width = 1 \linewidth]{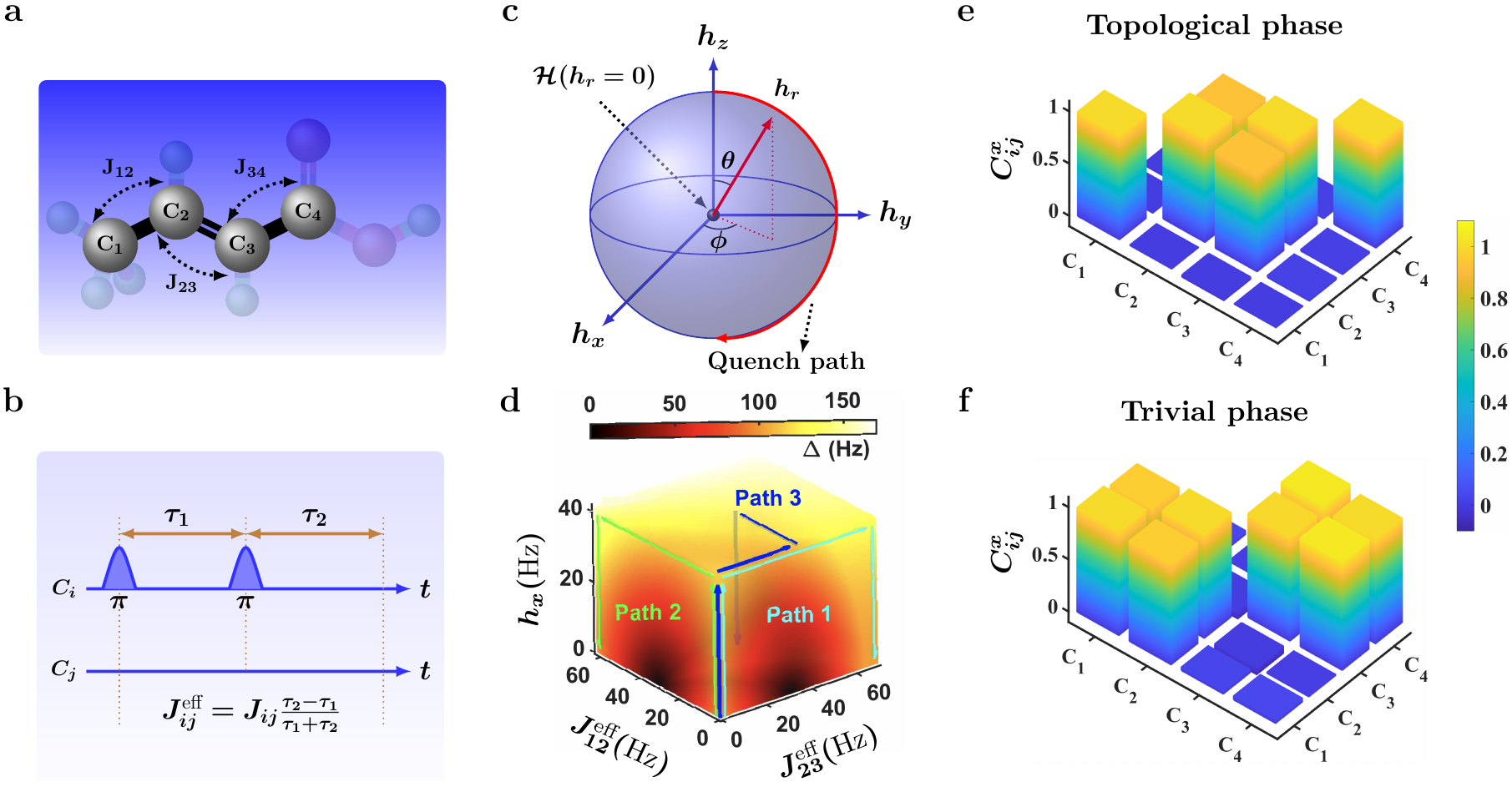}
\caption{\textbf{Experimental realization of the SSH model and measured spin-spin correlations for different phases.} \textbf{a,} Molecular structure of $^{13}$C-labeled trans-crotonic acid, where $\text{C}_1, \text{C}_2, \text{C}_3 \text{ and }\text{C}_4$ nuclear spins are used as a four-qubit quantum simulator to realize the SSH model. All hydrogen nuclei (green balls) are decoupled throughout the experiments. The nearest-neighbor couplings of $J_{12}=41.6$ Hz, $J_{23}=69.7$ Hz, and $J_{34}=72.2$ Hz are dominating. \textbf{b,} Refocusing scheme to create an effective coupling $J_{ij}^{\text{eff}}$, by a pair of selective $\pi$ pulses acting on $\text{C}_i$ and $\text{C}_j$. By adjusting $\tau_1$ and $ \tau_2$, we can implement two topologically distinct configurations of the SSH model. \textbf{c,} SSH model in a spherical space of external magnetic field $\vec{h}$.  When $h_r =0$, the origin point (small black ball) represents the pure Hamiltonian of SSH model, i.e., $\mathcal{H}(h_r=0) = \mathcal{H}_{\text{SSH}}^{\text{S}}$. The quench path (red curve) starts at the north pole and ramps along the $\phi = \pi/2$ meridian with a constant velocity $v_{\theta}$ until it reaches the south pole at the final time $t_f$. \textbf{d,} Adiabatic paths (labeled by $1, 2, 3$) for preparing different ground states,  corresponding to a topological phase, a trivial phase and an arbitrary phase for given $J_{12}^{\text{eff}}\approx J_{34}^{\text{eff}}, J_{23}^{\text{eff}}$, respectively. The colormap indicates the energy gap between the ground state and the first excited state. \textbf{e} and \textbf{f} show the observed spin-spin correlations along the $x$ direction of topological and trivial phases. } \label{fig: Fig1}
\end{figure}

We experimentally realized the 1D SSH model of the form of equation (\ref{eq: spinHamofSSH}) using the $^{13}$C-labeled trans-crotonic acid dissolved in d6-acetone, whose molecular structure is depicted in Fig. \ref{fig: Fig1}a. It consists of four carbon-13 nuclear spins (gray balls) and six hydrogen-1 nuclear spins (green balls).  The oxygen-16 nuclei (red balls, 99.76\% abundance) have no nuclear magnetic resonance (NMR) signals due to its spin number $I=0$.  All $^{1}$H nuclear spins were decoupled by the WALTZ-16 decoupling sequence throughout the experiments \cite{Shaka1983Improved}. The internal spin interactions of four carbon-13 nuclei ($I=1/2$) read as $\mathcal{H}_{\text{int}}=\frac{\pi}{2}\sum_{i=1}^3 J_{i, i+1}\sigma_i^z\sigma_{i+1}^z$. Here we only consider the nearest-neighbor couplings of $J_{12}=41.6$ Hz, $J_{23}=69.7$ Hz, and $J_{34}=72.2$ Hz, which are dominant compared to others. The chemical shifts and undesired $J$-couplings were refocused by spin-echo sandwiches \cite{SM2}.  One can apply a pair of selective $\pi$ pulses shown in Fig. \ref{fig: Fig1}b to create the effective $J$-couplings, $J_{ij}^{\text{eff}}=J_{ij}\frac{\tau_2-\tau_1}{\tau_1+\tau_2}$, which are tunable by changing the intervals of $\tau_1$ and $\tau_2$. The $\sigma_i^x\sigma_{i+1}^x$ ($\sigma_i^y\sigma_{i+1}^y$) terms can be generated via rotating the $\sigma_i^z\sigma_{i+1}^z$ terms by a pair of $\pi/2$ pulses along the $y$ ($x$) axis. Therefore, we realized the effective Hamiltonian of $\mathcal{H}_{\text{SSH}}^{\text{S, eff}}=-\frac{\pi}{2}\sum_{i=1}^3J_{i,i+1}^{\text{eff}}(\sigma_i^x\sigma_{i+1}^x + \sigma_i^y\sigma_{i+1}^y)$ \cite{SM2}, corresponding to the spin Hamiltonian of 1D SSH model, for example, which can be set to the topological configuration with $J_{12}^{\text{eff}}, J_{34}^{\text{eff}}\approx 0 \text{Hz}, J_{23}^{\text{eff}}=69.7 \text{Hz}$, and the trivial configuration with $J_{12}^{\text{eff}}, J_{34}^{\text{eff}}=41.6 \text{Hz}, J_{23}^{\text{eff}}\approx 0 \text{Hz}$, respectively.

To capture the central characteristics of two topologically distinct phases, we introduced an external magnetic field $\vec{h}=(h_r \text{sin}\theta \text{cos}\phi, h_r \text{sin}\theta \text{sin}\phi, h_r \text{cos}\theta)$ into the SSH model, as shown in Fig. \ref{fig: Fig1}c, where the origin point (small black ball) represents the pure Hamiltonian of SSH model, i.e., $\mathcal{H}(h_r=0) = \mathcal{H}_{\text{SSH}}^{\text{S}}$. The Hamiltonian in the external magnetic field is $\mathcal{H}_{\text{ext}}=-\sum_{i=1}^{2N} \vec{h}\cdot\vec{\sigma_i}$, which was readily achieved by applying a radio-frequency (RF) pulse \cite{SM2}. Now we can define the Chern number over a closed manifold surface of parameter space of $\vec{h}$ as
\begin{equation}
	\mathcal{C}h(h_r)=\frac{1}{2\pi}\oint_{|\vec{h}|=h_r} d S_{\theta\phi}\mathcal{F}_{\theta\phi}.
\end{equation}
Here $\mathcal{F}_{\theta\phi}\equiv\partial_{\theta}\mathcal{A}_{\phi}-\partial_{\phi}\mathcal{A}_{\theta}$ is the Berry curvature, and $\mathcal{A}_{\theta}\equiv i \langle\psi_g|\partial_{\theta}|\psi_g\rangle$ is the Berry connection. The $|\psi_g\rangle$ denotes the ground state of system. The topology of Hamiltonian (\ref{eq: spinHamofSSH}) can thus be characterized by $\mathcal{C}h(0) = \text{lim}_{h_r \rightarrow 0} \mathcal{C}h(h_r)$. We have $\mathcal{C}h(0)\neq 0$ and $\mathcal{C}h(0)=0$ for the topological and trivial phases respectively.

The ability to measure the Berry curvature is crucial for our experiment. Based on a method proposed in \cite{Gritsev2012Dynamical}, we measured the responses of the generalized force along the $\phi$-direction, $\mathcal{M}_{\phi}= -\langle\psi(t)|\partial_{\phi}\mathcal{H}|\psi(t)\rangle$. The $|\psi(t)\rangle$ is an instantaneous state when the external magnetic field evolves with a constant velocity $v_{\theta} = \pi/t_f$ along the quench path, as shown in the red curve of Fig. \ref{fig: Fig1}c, where $\phi=\pi/2$. The $\mathcal{M}_{\phi}$ at $\phi=\pi/2$ represents the sum of local spin magnetizations in time, because $\mathcal{M}_{\phi}|_{\phi=\pi/2} =-h_r \text{sin}\theta(t)\sum_{j=1}^4\langle\psi(t)|\sigma_j^x|\psi(t)\rangle$, with $\theta(t) = v_{\theta} t$. Then the Berry curvature can be extracted from the linear response of $\mathcal{M}_{\phi}$, according to
\begin{equation}\label{eq: linearResponse}
	\mathcal{M}_{\phi} =\text{const}+\mathcal{F}_{\theta\phi}v_{\theta}+\mathcal{O}(v_{\theta}^2).
\end{equation}
 If the $v_{\theta}$ is sufficiently slow,  the $\mathcal{O}(v_{\theta}^2)$ can be negligible. The constant term gives the value of $\mathcal{M}_{\phi}$ in the adiabatic limit. Note that the Hamiltonian (\ref{eq: spinHamofSSH}) has the rotation invariance, meaning that the Berry curvature is independent of $\phi$ \cite{SM2}. The Chern number is hence simplified into the integral with respect to $\theta$ from $0$ to $\pi$. 

For the measurement of Berry curvature, we begin with the initial ground states at the north pole of spherical parameter manifold of an external magnetic field. The initial states with different configurations of SSH model were adiabatically prepared along the three paths illustrated in Fig. \ref{fig: Fig1}d. The evolution time of adiabatic preparation obeys $T=\mathcal{O}(1/\Delta_{\text{min}}^2)$, where $\Delta_{\text{min}}$ denotes the minimum energy gap between the ground state and the first excited state. The auxiliary sweeps of $h_x$ were added to avoid the closed gap. The details on the preparation of topological phase, for example, can be found in supplementary materials. We performed the measurements of the spin-spin correlations along the $x$ direction, $C_{ij}^x = \langle\sigma_i^x\sigma_j^x\rangle -\langle\sigma_i^x\rangle\langle\sigma_j^x\rangle$, for different initial states using five independent readout experiments \cite{SM2}. Figures \ref{fig: Fig1}e and \ref{fig: Fig1}f show the intercell spin correlation with $C_{23}^{x} = 0.926$ for the topological phase, and intracell spin correlations with $C_{12}^x = 0.960, C_{34}^x = 1.042$ for the trivial phase, respectively, where the measured signals have been normalized using the initial ground states. In the topological phase, two edge spins without correlations can be in spin up or down states independently, which gives rise to the four-fold ground state degeneracy. 

\begin{figure}
\centering
\includegraphics[width = 1 \linewidth]{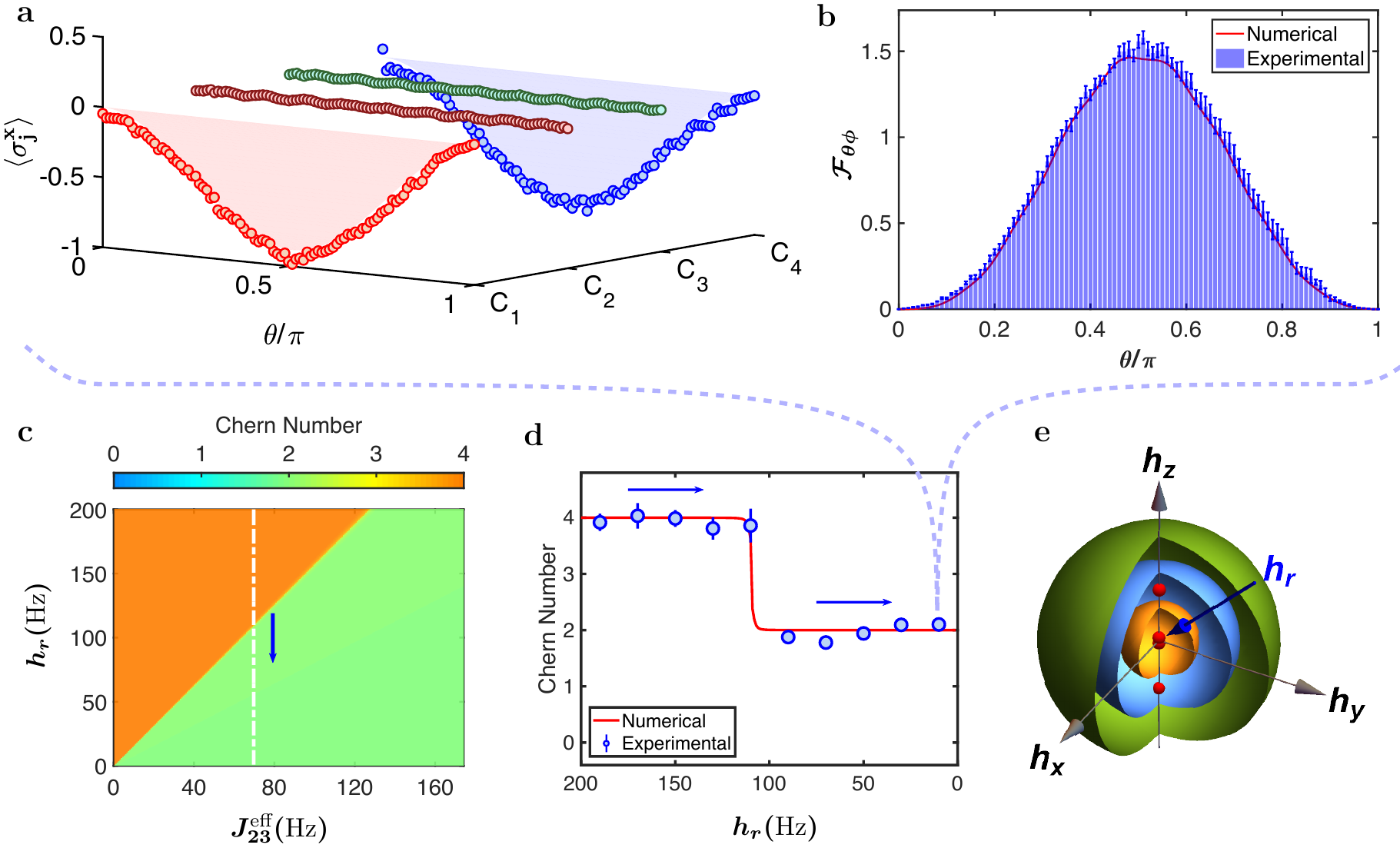}
\caption{\textbf{Observation of the topological phase.} The system was set to topological configuration, i.e., $J_{12}^{\text{eff}}, J_{34}^{\text{eff}}\approx 0 \text{Hz}$ and $J_{23}^{\text{eff}}=69.7 \text{Hz}$. \textbf{a,} Measured spin magnetizations $\langle\sigma_j^x\rangle$ for $j=1, 2, 3, 4$, which were generated along the quenching path shown in Fig. \ref{fig: Fig1}c. The initial ground state was adiabatically prepared at the north pole of spherical manifold ($h_r = 10$ Hz) through the path 1 (cyan lines) illustrated in Fig. \ref{fig: Fig1}d,  and each $\langle\sigma_j^x\rangle$ was measured at 101 time steps during the quenching evolution. The light shadings show the numerical results. \textbf{b,} Berry curvature extracted from the linear response of spin magnetizations. \textbf{c,} Calculated phase diagram showing the Chern number as a function of external magnetic field $h_r$ and effective coupling $J_{23}^{\text{eff}}$. The dotted white line at $J_{23}^{\text{eff}}=69.7$ Hz indicates the region explored experimentally in \textbf{d}. \textbf{d,} Measured Chern numbers in external magnetic fields, which were obtained from the integral of Berry curvature $\mathcal{F}_{\theta\phi}$. As $h_r \rightarrow 0$, the non-zero Chern number ($1.958\pm 0.070$) characterizes the topological phase directly. \textbf{e,} Schematic of measuring the Chern number that corresponds to the total number of degeneracy points (small red balls) enclosed in the spherical manifold. The degeneracy points at $h_r = 0$ indicate the non-zero Chern number of topological phase of the SSH model.} \label{fig: Fig2}
\end{figure}

\begin{figure}
\centering
\includegraphics[width = 1 \linewidth]{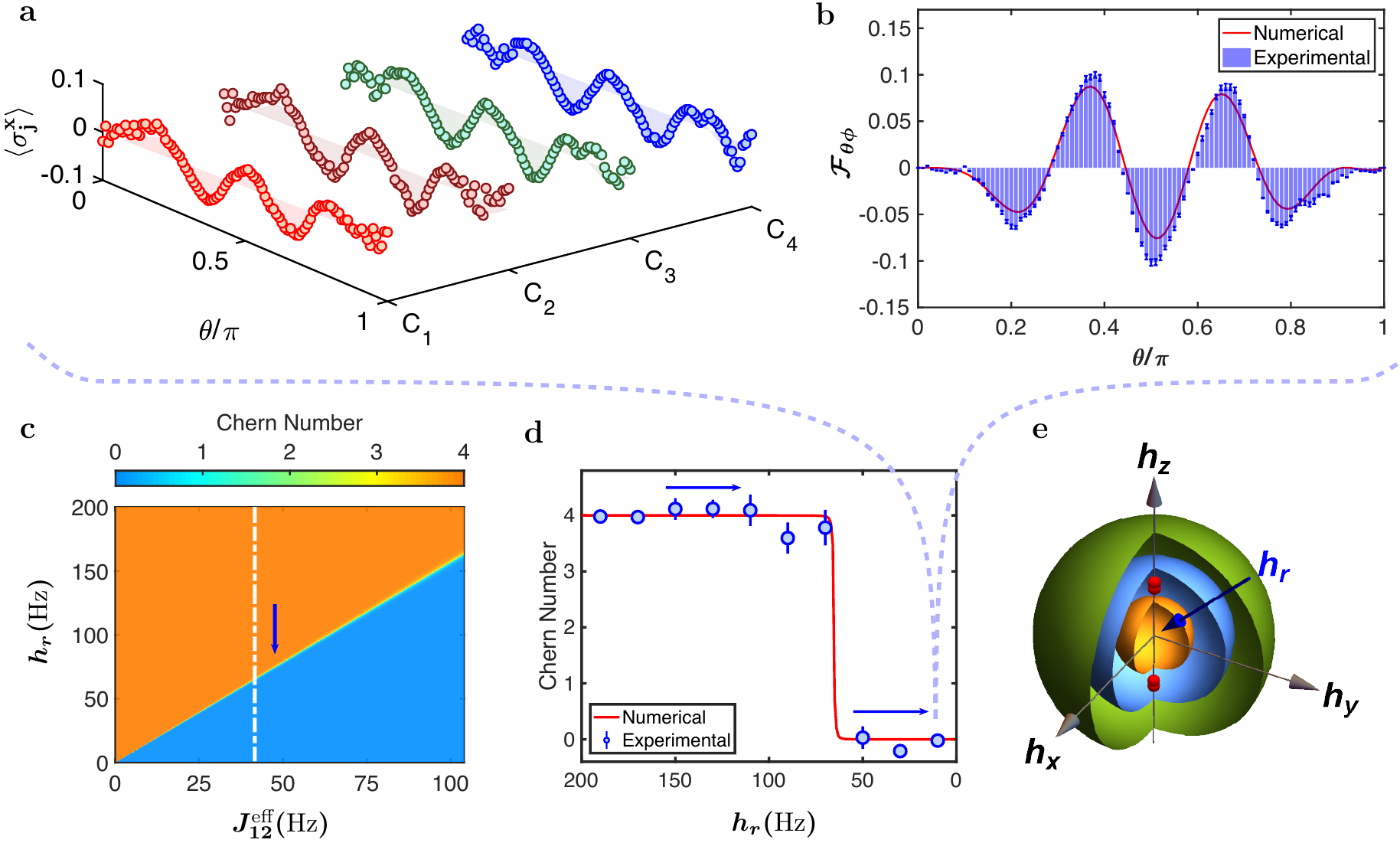}
\caption{\textbf{Observation of the trivial phase.} The system was set to trivial configuration, i.e., $J_{12}^{\text{eff}}, J_{34}^{\text{eff}} = 41.6 \text{Hz}$ and $J_{23}^{\text{eff}}\approx 0 \text{Hz}$. \textbf{a,} Measured spin magnetizations $\langle\sigma_j^x\rangle$ for $j=1, 2, 3, 4$, which were generated along the quenching path shown in Fig. \ref{fig: Fig1}c. The experiment began with an initial ground state preparing at the north pole of spherical manifold ($h_r = 10$ Hz) through the adiabatic evolution along the path 2 (green lines) illustrated in Fig. \ref{fig: Fig1}d,  and each $\langle\sigma_j^x\rangle$ was measured at 101 time steps during the quenching evolution. The light shadings show the numerical results. \textbf{b,} Berry curvature extracted from the linear response of spin magnetizations. \textbf{d,} Calculated phase diagram showing the Chern number as a function of external magnetic field $h_r$ and effective coupling $J_{12}^{\text{eff}}$. The dotted white line at $J_{12}^{\text{eff}}=41.6$ Hz indicates the region explored experimentally in \textbf{d}. \textbf{d,} Measured Chern numbers in external magnetic fields, which were obtained from the integral of Berry curvature $\mathcal{F}_{\theta\phi}$. As $h_r \rightarrow 0$, the near-zero Chern number ($-0.067\pm 0.063$) characterizes the trivial phase directly. \textbf{e,} Schematic of measuring the Chern number that corresponds to the total number of degeneracy points (small red balls) enclosed in the spherical manifold. There are no degeneracy points at $h_r = 0$, which indicates that it is the trivial phase of the SSH model.} \label{fig: Fig3}
\end{figure}

We now turn to observe two topologically distinct phases with the measured Chern number under different external magnetic fields. The system was set to the topological and trivial configurations of the SSH model. The phase diagrams of two configurations in the presence of external magnetic fields were plotted in Figs. \ref{fig: Fig2}c and \ref{fig: Fig3}c, showing the calculated Chern number as a function of external magnetic field $h_r$ and effective coupling $J_{23}^{\text{eff}}$ or $J_{12}^{\text{eff}}$.  For large $h_r$, both figures exhibit the orange regions with $\mathcal{C}h = 4$, corresponding to the spin-polarized (SP) state induced by the external magnetic fields \cite{SM2}. We experimentally explored the regions along the dotted white lines, with fixed $J_{23}^{\text{eff}}=69.7$ Hz (Fig. \ref{fig: Fig2}c) and $J_{12}^{\text{eff}}=41.6$ Hz (Fig. \ref{fig: Fig3}c). The main results of Chern number are shown in Figs. \ref{fig: Fig2}d and \ref{fig: Fig3}d, where each $\mathcal{C}h(h_r)$ was obtained by the integrals of the Berry curvatures, which were further extracted from the linear responses of the spin magnetizations to the quench velocity using equation (\ref{eq: linearResponse}). The quench time $t_f = 350$ ms taken in our experiments was slow enough, such that the high-order errors remain small \cite{SM2}. Figures \ref{fig: Fig2}a and \ref{fig: Fig2}b show the measured spin magnetizations and resulting Berry curvatures for $h_r = 10$ Hz in the topological configuration. Figures \ref{fig: Fig3}a and \ref{fig: Fig3}b show the similar results in the trivial configuration. Other relevant data for measuring the Chern numbers can be found in supplementary materials. 

The Chern number reveals the singularities of the system, counting the total number of degeneracy points enclosed in the spherical parameter manifold of $\vec{h}$, which may be viewed as the sources of magnetic monopoles from the Gauss' law \cite{SM2}. In Figs. \ref{fig: Fig2}c and \ref{fig: Fig3}c, we obtained $\bar{\mathcal{C}h} = 1.958 \pm 0.070$ for the topological phase, and $\bar{\mathcal{C}h} = -0.067 \pm 0.063$ for the trivial phase, respectively, as the $h_r$ trends to zero. Note that each two-fold degenerate ground state forms a degeneracy point. The emergence of two degeneracy points (small red balls) at $h_r = 0$ determines the four-edge states, featuring an SPT phase, according to the bulk-edge correspondence (Fig. \ref{fig: Fig2}e). But the absence of a degeneracy point at $h_r = 0$ indicates that it is a trivial phase (Fig. \ref{fig: Fig3}e). We also demonstrated the different robustness of degeneracy points against the small perturbations (see supplementary materials). These signatures clearly distinguish the two topologically different phases of the SSH model. 

\begin{figure}
\centering
\includegraphics[width = 0.5 \linewidth]{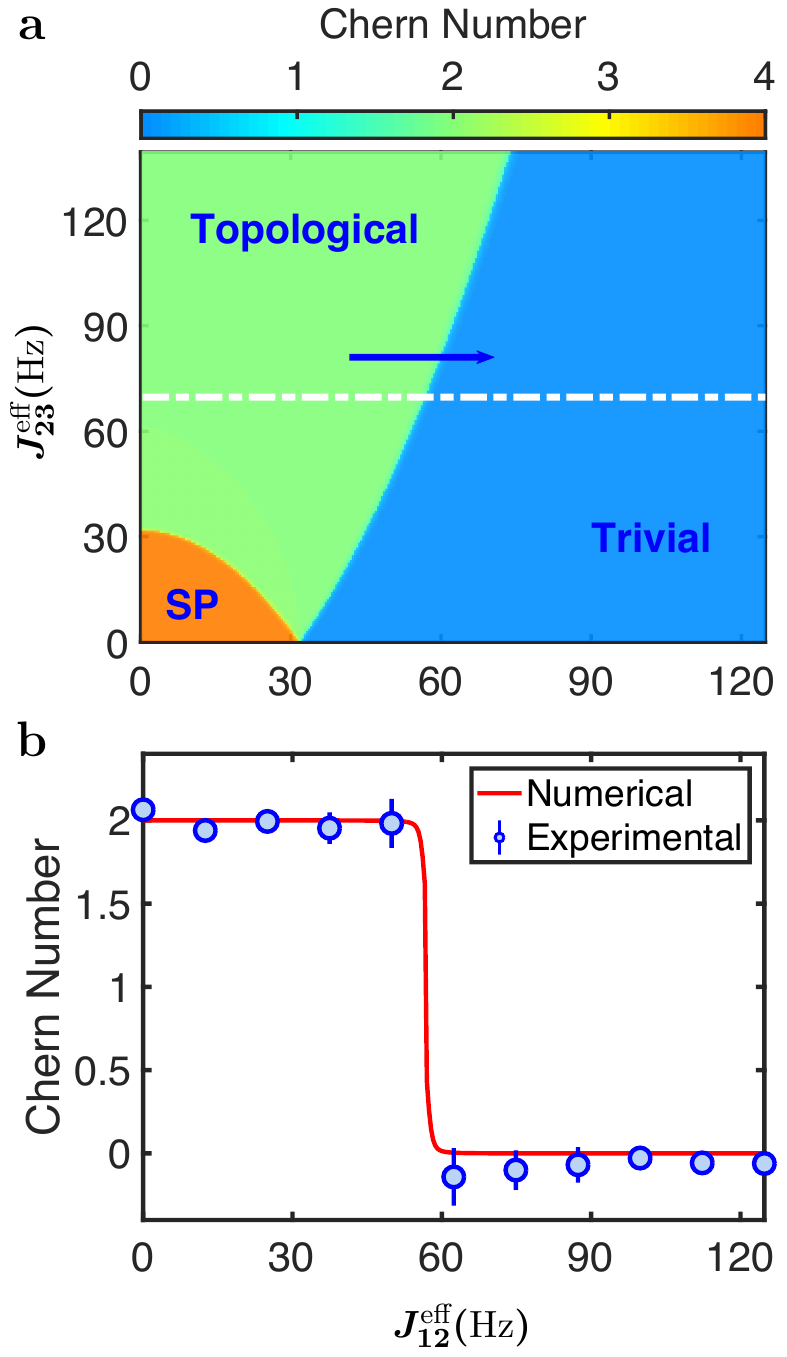}
\caption{\textbf{Probing the topological transition.} \textbf{a,} Calculated phase diagram showing the Chern number as a function of effective couplings $J_{12}^{\text{eff}}$ (or $J_{34}^{\text{eff}}$, $J_{34}^{\text{eff}}=J_{12}^{\text{eff}}$) and $J_{23}^{\text{eff}}$, with a fixed external magnetic field $h_r = 50$ Hz. The light green, light blue and orange regions denote the topological, trivial and spin-polarized (SP) phases, respectively. The dotted white line at $J_{23}^{\text{eff}}=69.7$ Hz indicates the region explored experimentally in \textbf{b}. \textbf{b,} Measured Chern number as a function of $J_{12}^{\text{eff}}$. The solid red line shows the numerical simulation and experimental data are obtained by integrating the Berry curvature (see supplementary materials).}\label{fig: Fig4}
\end{figure}

We further probed the interaction-induced transition between two topologically distinct phases, where the external magnetic field strength was set to the small value of $h_r = 50$ Hz. The calculated phase diagram exhibits three different regions of light green, light blue and orange, corresponding to the topological phase with Chern number two, the trivial phase with Chern number zero, and the SP phase with Chern number four (Fig. \ref{fig: Fig4}a). The area of SP phase depends on the field strength $h_r$, and in the absence of interactions, its Chern number grows linearly with the increase of system size, due to the introduced external magnetic field itself (see supplementary materials). We experimentally explored the region along the dotted white line with fixed $J_{23}^{\text{eff}} = 69.7$ Hz. By adjusting $J_{12}^{\text{eff}}$ and $J_{34}^{\text{eff}}$, we observed a sharp transition from the nontrivial Chern number ($\bar{\mathcal{C}h} = 1.986\pm 0.048$) to the vanishing value ($\bar{\mathcal{C}h} = -0.076\pm 0.040$) (Fig. \ref{fig: Fig4}b), as expected from numerical calculations, indicating the change in topology. 

In conclusion, we have presented the experimental realization of a 1D SSH model with interacting nuclear spins, and implemented a general method for the real-space measurements of many-body Chern numbers through the linear responses to the external magnetic fields. As the introduced field strength decreases to zero, we have observed an interacting SPT phase with the nonzero Chern number, that determines the number of the degenerate ground states and also the nontrivial edge states according to the bulk-edge correspondence, and a trivial phase with zero Chern number, which has only a single ground state. In fact, it is not necessary to measure the Chern number in the zero magnetic field limit, because which is invariant in weak fields for many cases \cite{SM2}. The method for experimentally extracting the Chern number is also robust against the noises of applied fields, regardless of the shapes of the closed surface \cite{SM2}. Using the measured Chern number, we have further probed the interaction-induced transition between two topologically distinct phases under a fixed small external field.   

Our results provide the direct and central characterization of an interacting SPT phase in real space. The realization techniques of the SSH model (e.g., refocusing scheme and quenching control) can be extended to larger nuclear spin systems, such as the $^{13}$C-labeled polyacetylene, for further exploring the more interesting phenomena, though the scalability of NMR platform in quantum computing might be limited. Besides, the dynamical method of measuring the Chern number can be generalized to other platforms and momentum spaces, and may provide a promising route for topological characterization of many-body states in higher dimensional systems, such as the topologically ordered states with long-range quantum entanglement.


\begin{addendum}

 \item This work was supported by the National Natural Science Foundation (Grants No. 11805008, No. 12074206, No. 12104213, No. 11734002, No. 12075110, and No. 11875159), Guangdong Provincial Key Laboratory (Grant No. 2019B121203005), Guangdong Innovative and Entrepreneurial Research Team Program (2019ZT08C044), and Science, Technology and Innovation Commission of Shenzhen Municipality (KQTD20190929173815000 and JCYJ20200109140803865).

 \item[Author Contributions] Z. L. proposed the idea, designed and performed the experiment. Z. L. analyzed the data and wrote the manuscript. X. N. prepared the sample. Z. L. and W. Z. developed the theoretical framework.  Z. L. and D. L. revised the manuscript and supervised the work.  All authors discussed the results and contributed to the writing of the manuscript. 
 \item[Competing Interests] The authors declare that they have no competing interests.
 \item[Correspondence] Correspondence and requests for materials
should be addressed to Z. L. (luozhih5@mail.sysu.edu.cn) or D. L. (ludw@sustech.edu.cn).
\end{addendum}

\clearpage

\clearpage
\setcounter{figure}{0}
\setcounter{equation}{0}
\setcounter{table}{0}
\setcounter{section}{0}
\setcounter{subsection}{0}
\setcounter{page}{1}
\renewcommand{\thefigure}{S\arabic{figure}}
\renewcommand{\theequation}{S\arabic{equation}}
\renewcommand{\thetable}{S\arabic{table}}
\renewcommand{\thesection}{S\arabic{section}}

\begin{center}
{\bf{SUPPLEMENTARY INFORMATION}\\
\bf{for}\\
\bf{Observation of a symmetry-protected topological phase in external magnetic fields}\\}
\author{Zhihuang Luo${^{1}}^{*}$, Wenzhao Zhang$^{2}$, Xinfang Nie$^{3}$, \& Dawei Lu${^{3}}^{*}$}
\end{center}

\begin{affiliations}
 \item Guangdong Provincial Key Laboratory of Quantum Metrology and Sensing \& School of Physics and Astronomy, Sun Yat-Sen University (Zhuhai Campus), Zhuhai 519082, China.
 \item Department of Physics, Ningbo University, Ningbo 315211, China.
 \item Shenzhen Institute for Quantum Science and Engineering and Department of Physics, Southern University of Science and Technology, Shenzhen 518055, China.
\end{affiliations}

\tableofcontents
\clearpage

\section{SSH model in external magnetic fields}

\subsection{SSH model}

\begin{figure}
\centering
\includegraphics[width = 0.95 \linewidth]{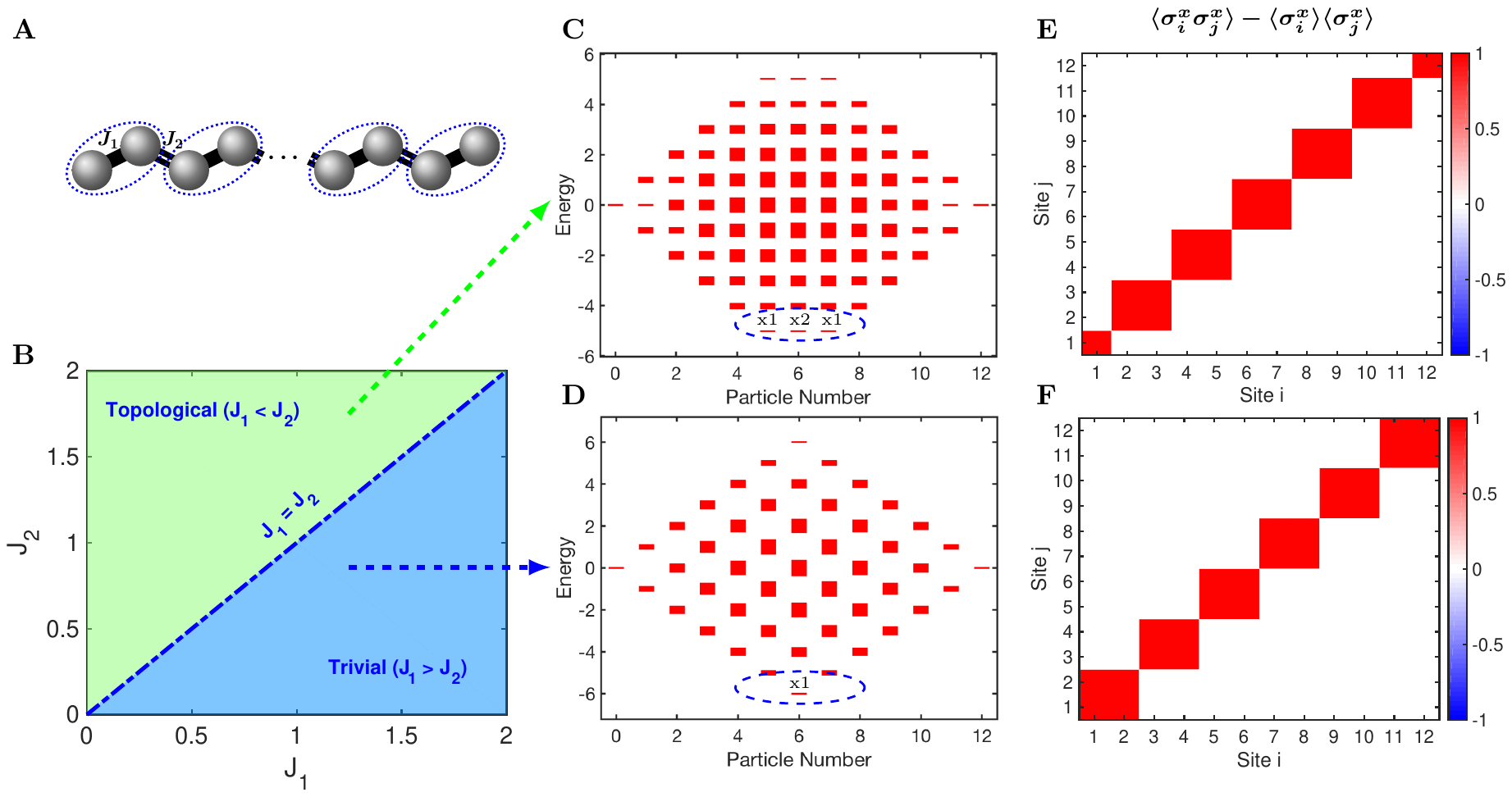}
\caption{(A) Geometry of the SSH model in a one-dimensional lattice with open boundary conditions. Each two sites linked by thick lines is grouped into a unit cell (circled by a dotted line). The hopping amplitudes are staggered, i.e., the inracell hoppings $J_1$ (thick lines) are different from the intercell hoppings $J_2$ (doubled thin lines). (B)  Phase diagram of the SSH model. There are two distinct phases: topological one (green) when $J_1 < J_2$ and trivial one (blue) when $J_1 > J_2$. The boundary of $J_1 = J_2$  separates these phases. (C and D) Full energy spectrum of the spin chain of 12 sites in the topological and the trivial configurations for different particle numbers. The topological configuration exhibits a fourfold degeneracy involving five, six (twofold degenerate), and seven particles, whereas the trivial chain exhibits a single gapped ground state with six particles. (E and F) Calculated spin-spin correlations along the x direction for the topological and trivial phases. }\label{figs: modelPhase}
\end{figure}

We consider the Su-Schrieffer-Heeger (SSH) model in a one-dimensional (1D) lattice with $N = L/2$ unit cells \cite{Su1979Solitons, Asboth2016A}, each unit cell consisting of 2 fermions, as illustrated in Fig. \ref{figs: modelPhase}A. Its Hamiltonian reads
\begin{equation}\label{eqs: HamofSSH}
	\mathcal{H}_{\text{SSH}}^{\text{F}}=-J_1\sum_{i=1}^N c_{2i-1}^{\dag} c_{2i} - J_2\sum_{i=1}^{N-1} c_{2i}^{\dag} c_{2i+1} + h.c,
\end{equation}
where $c_i^{\dag}$ and $c_i$ represent the fermionic creation and annihilation operators. The hopping amplitudes are staggered, meaning that the intracell hoppings $J_1$ are different from the intercell hoppings $J_2$. Depending on the ratio of $J_1$ and $J_2$, the SSH model exhibits two topologically distinct phases, shown In Fig. \ref{figs: modelPhase}B. If $J_1 < J_2$, the system is in the symmetry-protected topological (SPT) phase with four-fold degenerate ground states (Fig. \ref{figs: modelPhase}C); if $J_1 > J_2$,  the system is in the trivial phase with a single ground state (Fig. \ref{figs: modelPhase}D). The SSH model is a paradigmatic example for an SPT phase. The existence of SPT phase is relevant to two important symmetries: particle number conversation and chiral symmetry that protects the topological phase. 

Using the following Jordan-Wigner transformation
\begin{equation}
       c_j=\prod_{i=1}^{j-1}\sigma_{i}^z\sigma_j^{+},\quad  c_j^{\dag}=\prod_{i=1}^{j-1}\sigma_{i}^z\sigma_j^{-},
\end{equation}
the SSH Hamiltonian (\ref{eqs: HamofSSH}) can be rewritten in the form of spin-1/2 operators as
\begin{equation}\label{eqs: Ham2ofSSH}
    \mathcal{H}_{\text{SSH}}^{\text{S}}=-J_1\sum_{i=1}^N (\sigma_{2i-1}^x\sigma_{2i}^x +\sigma_{2i-1}^y\sigma_{2i}^y) 
    - J_2 \sum_{i=1}^{N-1} (\sigma_{2i}^x\sigma_{2i+1}^x +\sigma_{2i}^y\sigma_{2i+1}^y),
\end{equation}
where $\sigma_i^{\alpha} (\alpha=x, y, z)$ represent the Pauli matrices. This is the spin Hamiltonian of SSH model considered in our experiment. Using the corresponding many-body ground states of Eq. (\ref{eqs: Ham2ofSSH}), the numerical calculations of the spin-spin correlations along $x$ direction, $C_{ij}^x = \langle\sigma_i^x\sigma_j^x\rangle -\langle\sigma_i^x\rangle\langle\sigma_j^x\rangle$, are performed on a chain of 12 sites. We obtain $C_{2i,2i+1}^x =1$ for the topological phase, and $C_{2i-1,2i}^x =1$ for the trivial phase, showing the inter- and intracell correlations between spins in Figs. \ref{figs: modelPhase}E and \ref{figs: modelPhase}F, respectively. It could be found that for the topological phase, there are two independent edge spins, different from the correlated bulk spins. The appearance of edge states at the boundary spins gives rise to the four-fold ground state degeneracy. 

\subsection{Many-body Chern number}\label{subsec: ChernNumber}

To describe the central properties of SPT phase using the topological invariants, we shall work with the following generic Hamiltonian by introducing an external magnetic field of $\vec{h}=(h_r \text{sin}\theta \text{cos}\phi, h_r \text{sin}\theta \text{sin}\phi, h_r \text{cos}\theta)$ into the spin form of SSH model
\begin{equation}\label{eqs: totHam}
	\mathcal{H}(\vec{h}) = -\sum_{i=1}^{L} \vec{h}\cdot\vec{\sigma_i} + \mathcal{H}_{\text{SSH}}^{\text{S}}.
\end{equation}
It allows for defining the Berry curvature over the spherical parameter surface of $\vec{h}$, and the further relevant Chern number obtained by the integral of the Berry curvature. The Chern number manifests the singularities of system enclosed in the spherical manifold, and can thus be used to capture the signature of topology. If the field strength decreases to zero,  the nonzero value of Chern number reveals the existence of a nontrivial phase in Eq. (\ref{eqs: Ham2ofSSH}); on the contrary, it is a trivial phase. Note that the Hamiltonian (\ref{eqs: totHam}) has $U(1)$ invariance, i.e.,
\begin{equation}
	\mathcal{H}(h_r, \theta, \phi)= R_{\text{tot}}^z(\phi)\mathcal{H}(h_r, \theta, 0) R_{\text{tot}}^z(-\phi),
\end{equation}
with $R_{\text{tot}}^z=\prod_{i=1}^{L} e^{-i\phi\sigma_i^z/2}$. Therefore, the Berry curvature is independent of $\phi$, and the Chern number can be simplified into the integral of Berry curvature with respect to the $\theta$ from 0 to $\pi$. When $\theta=0$ and $\pi$, $\sum_{i=1}^{L} \sigma_i^z$ is a conversation quantity, which enables the points of ground state degeneracy locating on the $z$ axis of $\vec{h}$. The degeneracy points act as the sources of magnetic monopoles, and contribute the nonzero terms to the integral of Berry curvature, which can be seen more intuitively from the full expression of  Berry curvature
\begin{equation}\label{eqs: BerryCurvature}
	\mathcal{F}_{\theta\phi}(h_r) = i \sum_{n\neq 0}\frac{\langle\psi_0|\partial_{\theta}\mathcal{H}|\psi_n\rangle\langle\psi_n|\partial_{\phi}\mathcal{H}|\psi_0\rangle - \langle\psi_0|\partial_{\phi}\mathcal{H}|\psi_n\rangle\langle\psi_n|\partial_{\theta}\mathcal{H}|\psi_0\rangle}{(E_n - E_0)^2},
\end{equation}
using the definitions of $\mathcal{F}_{\theta\phi}$ and $\mathcal{A}_{\theta}$ in the main text. Here $E_n$ and $|\psi_n\rangle$ denote the $n_{\text{th}}$ eigenvalue and its corresponding eigenstate of Hamiltonian (\ref{eqs: totHam}). It clearly shows that the $E_n = E_0$ for $n\neq 0$ are the singular points of the above expression, corresponding to the ground state degeneracy. As the analogy with the Gauss's law, the Chern number in the integral form of  Eq. (\ref{eqs: BerryCurvature}) determines the number of degeneracy points, and also the number of edge states according to the bulk-edge correspondence. 

\subsection{Robustness against the external magnetic field}

\begin{figure}
\centering
\includegraphics[width = 0.9 \linewidth]{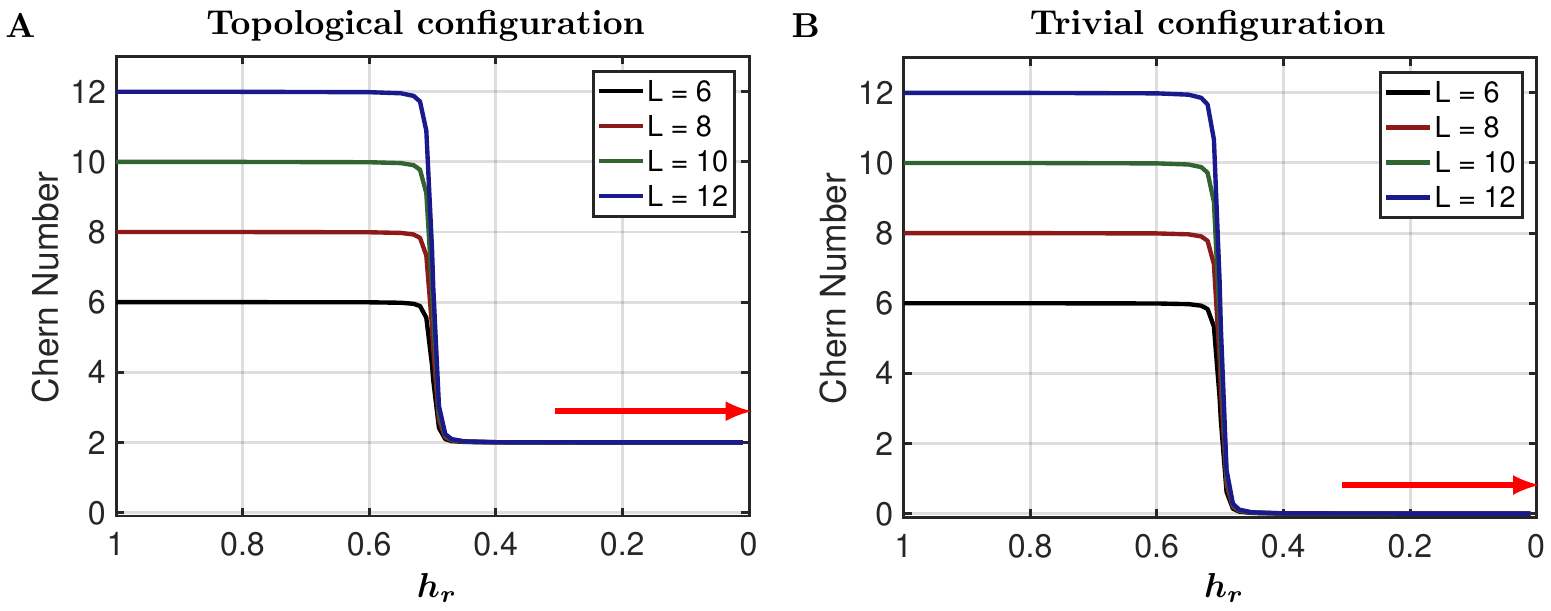}
\caption{Calculated Chern numbers as a function of the field strength $h_r$ for the topological configuration (A) and the trivial configuration (B), respectively. As the $h_r$ trends to zero, the Chern numbers are invariant with the value of two for the topological phase, and zero for the trivial phase, regardless of the system sizes. }\label{figs: ChernNumbervshrnCell}
\end{figure}

\begin{figure}
\centering
\includegraphics[width = 0.8 \linewidth]{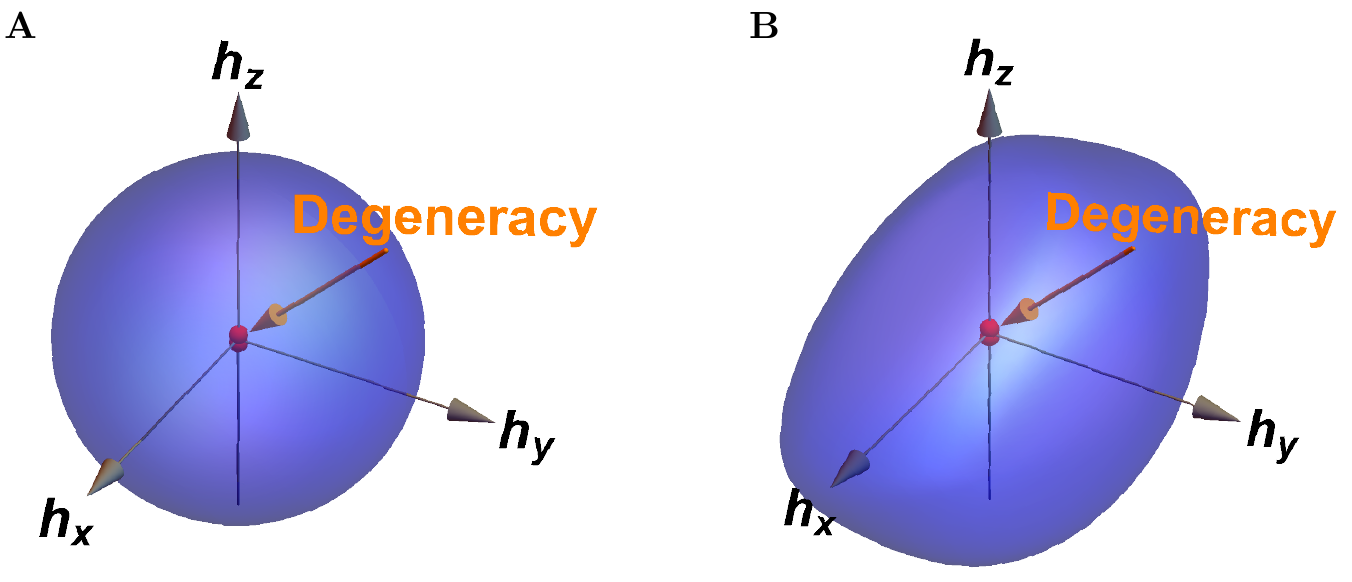}
\caption{The introduced external magnetic field is spherical (A) and arbitrary (B) closed surfaces. The small red balls enclosed in the closed surfaces represent the degeneracy points of Hamiltonian (\ref{eqs: Ham2ofSSH}) ($J_1 < J_2 $) without the external field, which act as the sources of magnetic monopoles.}\label{figs: fieldShape}
\end{figure}

By the exact diagonalization of Hamiltonians (\ref{eqs: totHam}) with $L = 6, 8, 10, 12$ and the integrals of Eq. (\ref{eqs: BerryCurvature}), we obtained the same numerical value of Chern number two for the topological phase when the $h_r$ trends to zero, as shown in Fig. \ref{figs: ChernNumbervshrnCell}A. It determines the four-fold degenerate edge states of SPT phase. For the trivial phase with a single ground state, the numerical value of Chern number is zero (Fig. \ref{figs: ChernNumbervshrnCell}B). It could be seen that the Chern number is invariant in weak fields, as it is in the zero magnetic field limit. Therefore, we can only measure the many-body Chern number in a weak field, which is feasible for many systems, and more efficient experimentally for extracting the topological features. The method for the measurement of Chern number is also robust against the noises of applied fields, because the value of Chern number is related to the distribution of magnetic monopoles, irrespectively of the shapes of a closed surface (see Figures. \ref{figs: fieldShape}A and \ref{figs: fieldShape}B). 

\section{Experimental realization}\label{sec: experimentalRealization}

\subsection{Setup and sample} 

\begin{figure}
\centering
\includegraphics[width = 0.85 \linewidth]{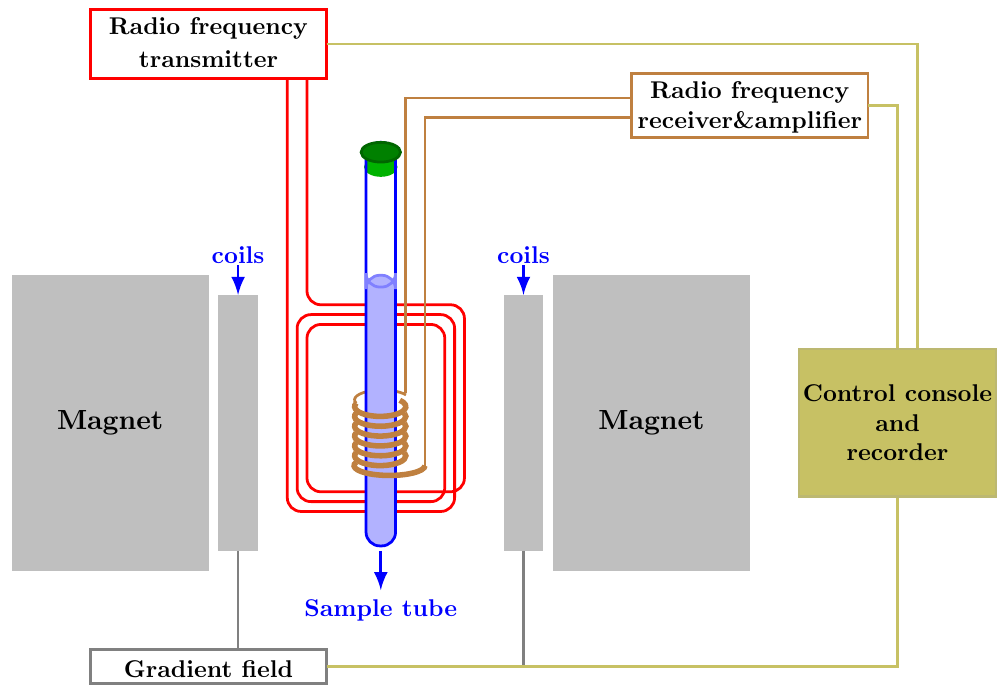}
\caption{ Schematic of an NMR apparatus, including a large superconducting magnet, the RF transmitter, receiver $\&$ amplifier, the control console and recorder, the gradient field coils, and the sample held in a glass tube. The superconducting magnet provides a static magnetic field $B_0 = 14.1$ Tesla along the $z$ axis. }\label{figs: NMRapparatus}
\end{figure}

The experiment was carried out on a Bruker DRX 600 MHz spectrometer at temperature 300 K. The spectrometer is constructed from a large superconducting magnet, the radio frequency (RF) section, gradient field coils, and the sample held in a glass tube, as shown in Fig. \ref{figs: NMRapparatus}. The superconducting magnet provides a static magnetic field along the $z$ axis, with the magnitude of $B_0=14.1$ Tesla, which is extremely homogeneous within at least 1 part in $10^9$. The RF section consists of the transmitter and receiver, being that part of the spectrometer used for generating and detecting the RF signals. The gradient field along the $z$ axis generated by the gradient field coils can be used to average out the terms of zero quantum coherence of density matrix. When the sample is placed in the magnetic field of $B_0$, the NMR signal is produced by excitation of the nuclei (such as $^1\text{H}, ^{13}\text{C}$) of sample molecule with resonant RF fields, and is detected with RF receivers. At the beginning of the experiment, the nuclear spins are thermalized to the equilibrium state by waiting a long period, which requires several minutes for well-prepared liquid samples. Under the external control, RF pulses are then applied to implement the desired unitary transformation on the initial state of spins. The final state is measured from the peaks of frequency spectrum obtained by the Fourier transformation of NMR signal that is generated via the free induction decay.

\begin{figure}
\centering
\includegraphics[width = 0.75 \linewidth]{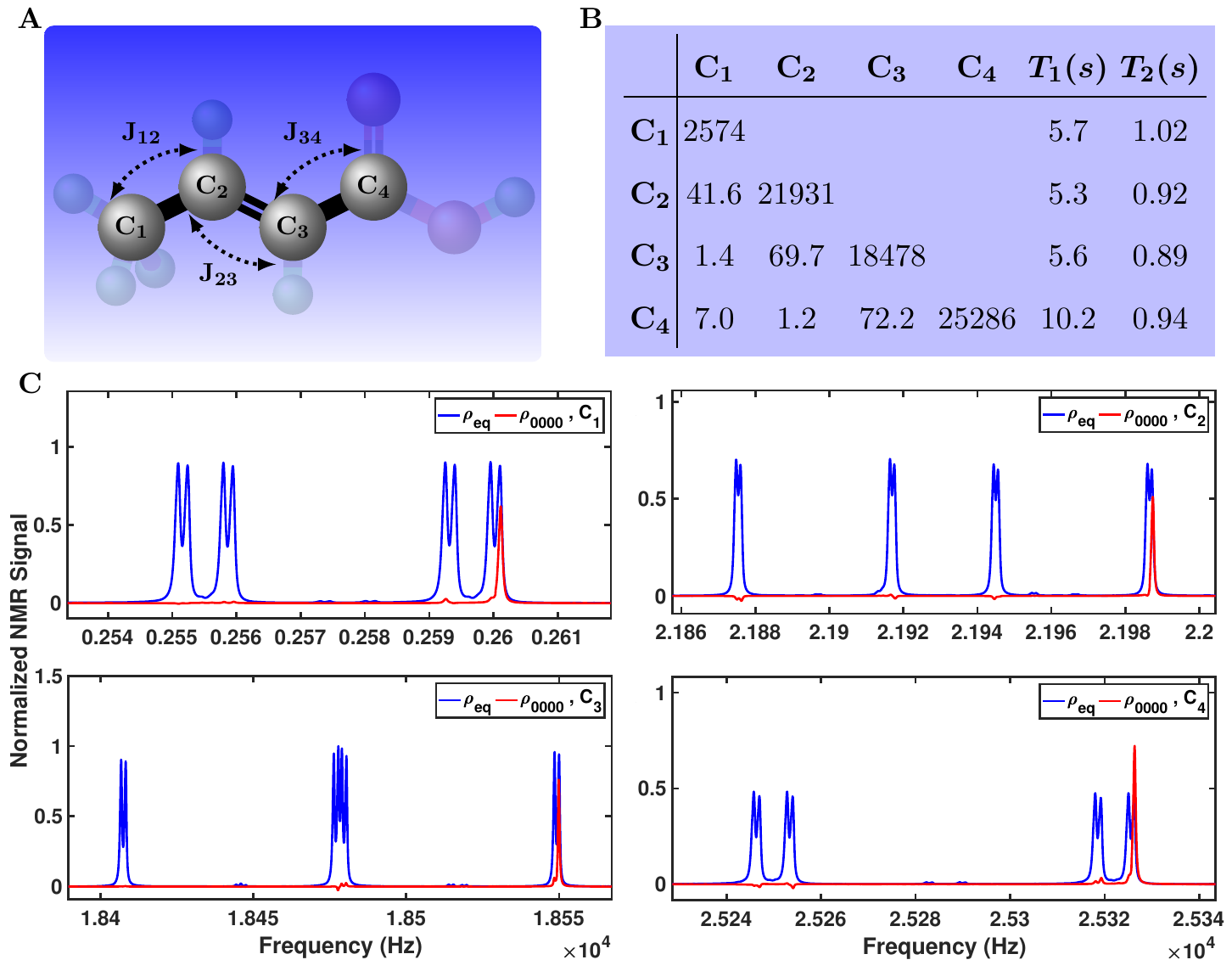}
\caption{ (A) Molecular structure of $^{13}$C-labeled trans-crotonic acid, where $\text{C}_1, \text{C}_2, \text{C}_3 \text{ and }\text{C}_4$ nuclear spins are used as a four-qubits quantum simulator. All hydrogen nuclei (green balls) are decoupled throughout the experiments. (B) Its Hamiltonian parameters, where diagonal and off-diagonal elements represent the chemical shifts and J-coupling constants (in Hz), respectively. The $T_1$ and $T_2$ are spin-lattice relaxation times and spin-spin relaxation times (in seconds).  (C) Experimental carbon spectra of thermal equilibrium state $\rho_{eq}$ (blue) and pseudo-pure state $\rho_{0000}$ (red). }\label{figs: CrotonicAcid}
\end{figure}

The sample employed in our experiments is the molecule of $^{13}$C-labeled trans-crotonic acid dissolved in d6-acetone, shown in Fig. \ref{figs: CrotonicAcid}A. It consists of four carbon nuclei with spin-1/2, which are used as our physical system. The natural Hamiltonian of this system in the rotating frame reads
\begin{equation}\label{eqs: NMRHamiltonian}
\mathcal{H}_{\text{NMR}}=\sum_{i=1}^{4}\frac{\omega _{i}}{2}\sigma_i^z+\sum_{i<j,=1}^{4}\frac{\pi J_{ij}}{2}\sigma_i^z\sigma_j^z,
\end{equation}
where $\omega _{i}$ represents the chemical shift of spin $i$, and $J_{ij}$ the coupling constant between spins $i$ and $j$. The relevant Hamiltonian parameters can be found in Fig. \ref{figs: CrotonicAcid}B. It shows that the nearest neighbor couplings are dominating compared to others. The other interactions with the $^1$ H spins were decoupled by the WALTZ-16 decoupling sequence throughout the experiments.

Figure \ref{figs: CrotonicAcid}C illustrates the $^{13}$C spectra of the thermal equilibrium state $\rho_{\text{eq}}$ and the pseudo-pure state (PPS) $\rho_{0000}$, which was prepared from the $\rho_{\text{eq}}$ by three line-selective shaped pulses. The first shaped pulse with length of 80 ms was applied to selectively excite the energy levels and make the populations equal of all levels except for $0000$-level. It can be written as the product of a sequence of line-selective pulses, i.e.,
\begin{equation}\label{}
    U_1=\prod_{n=1}^{14}e^{-i\theta_n\hat{I}_x^{(n_1,n_2)}},
\end{equation}
where $\hat{I}_x^{(n_1,n_2)}$ is the single-transition operator between energy levels $n_1$ and $n_2$. The flipping angles $\{\theta_n\}_{n=1,2,\cdots,14}$ were optimized by numerical search procedure, with appropriate values that satisfies the following condition: $\text{diag}[U_1\rho_{\text{eq}}U_1^{\dag}]=\text{diag}[\rho_{0000}]$.
The other shaped pulse with length of 24 ms were applied to remove zero quantum coherence that cannot be averaged out by $z$-direction gradient fields. They consists of the controlled-NOT gates between qubits $n$ and $m$, denoted as $\text{CNOT}_{nm}$. For example, the second and third shaped pulses of $U_2=\text{CNOT}_{12}$ and $U_3=\text{CNOT}_{34}$ were chosen in our experiment. The intensity ratios for the rightmost peak in PPS $\rho_{0000}$ and the reference in the equilibrium spectrum are $0.704, 0.782, 0.807, 1.605$ for $\text{C}_1, \text{C}_2, \text{C}_3, \text{C}_4$, respectively. The signal loss are caused by the gradient fields and the relaxation effect with different spin-lattice relaxation time $T_1$ for different carbon nuclei. The $T_1$ of $C_4$ is longer than other carbon nuclei.

\subsection{Realization of the SSH model}

\begin{figure}
\centering
	\includegraphics[width = 0.5 \linewidth]{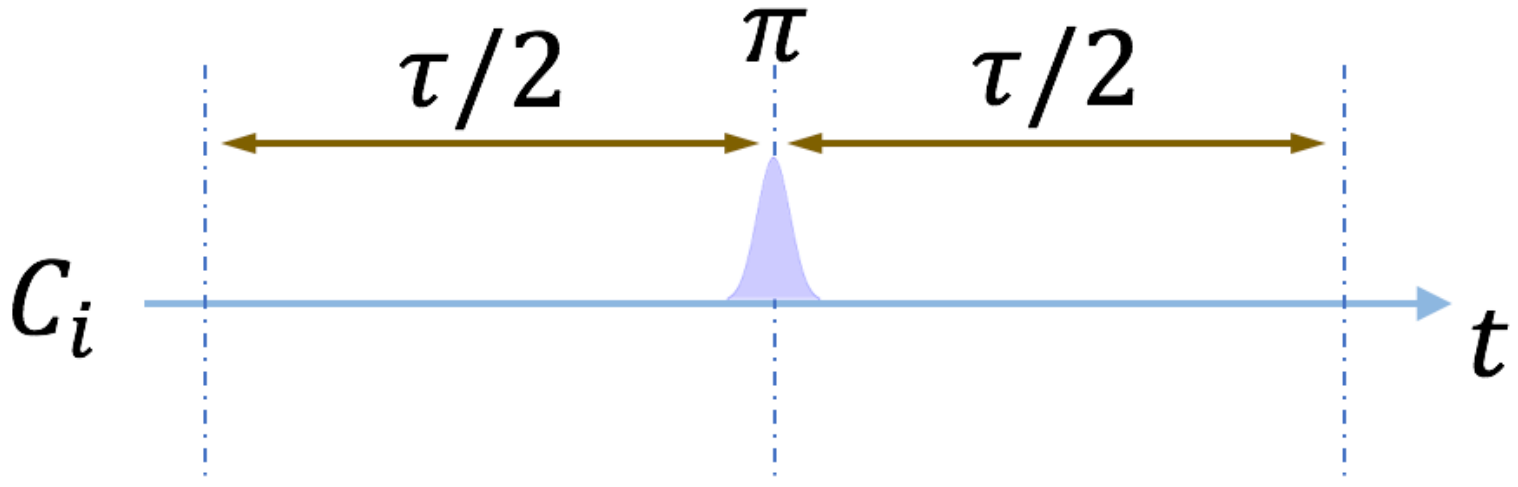}
	\caption{A spin echo sandwich of duration $\tau$. It consists of two intervals of free evolution, both of duration $\tau/2$, ``sandwiching'' a line-selective pulse of flip angle $\pi$.}\label{figs: refocusingPulse}
\end{figure}

Our realization of the SSH model is performed on the four nuclear spins of the trans-crotonic acid. In Eq. (\ref{eqs: NMRHamiltonian}), the terms of the chemical shifts $\omega_i$ and undesired couplings $J_{ij}$ for $j\neq i$ can be refocused by spin echo sandwiches, as shown in Fig. \ref{figs: refocusingPulse}. We get
\begin{equation}\label{eqs: spinEcho}
	U_{\text{free}}(\tau/2)R_i^x(\pi)U_{\text{free}}(\tau/2) = \text{exp}\left [-i\left (\sum_{j\neq i}^4\frac{\omega_j}{2}\sigma_j^z + \sum_{j,k \neq i}^4\frac{\pi J_{jk}}{2}\sigma_j^z\sigma_k^z \right ) \tau \right ] R_i^x(\pi),
\end{equation}
where $U_{\text{free}} (\tau/2)\equiv \text{exp}(-i\mathcal{H}_{\text{NMR}}\tau/2)$ represents the free evolution of the NMR Hamiltonian (\ref{eqs: NMRHamiltonian}), and $R_i^x(\pi)\equiv \text{exp}(-i\pi\sigma_i^x/2)$ stands for a selective $\pi$ pulse acting on spin $i$ along the $x$ axis. From the right-hand side of Eq. (\ref{eqs: spinEcho}), it could be found that the $\omega_i$ and $J_{ij}$ for $j\neq i$ terms are removed. Using the spin echo sandwiches, we keep the dominate couplings of $J_{12}=41.6$ Hz, $J_{23}=69.7$ Hz, and $J_{34}=72.2$ Hz, compared to others ($J_{13}=1.4$ Hz, $J_{14}=7.0$ Hz, $J_{24}=1.2$ Hz). So the corresponding spin interactions we reached here is 
\begin{equation}\label{eq:NMRinteraction}
    \mathcal{H}_{\text{int}}=\frac{\pi}{2}\sum_{i=1}^3 J_{i, i+1}\sigma_i^z\sigma_{i+1}^z,
\end{equation}
only with the nearest neighbor couplings. One can further apply the pulse sequence shown in Fig. 1B of main text to create the effective couplings $J_{ij}^{\text{eff}}$, where using a pair of selective $\pi$ pulses acting on $\text{C}_i$ and $\text{C}_j$, we have
\begin{equation}
	U_{ij}(\tau_2)R_i^x(\pi)U_{ij}(\tau_1)R_i^x(\pi) = \text{exp}\left [ -i\frac{\pi J_{ij}^{\text{eff}}}{2}\sigma_i^z\sigma_j^z (\tau_1+\tau_2)\right ]
\end{equation}
with $J_{ij}^{\text{eff}} = J_{ij}\frac{\tau_2-\tau_1}{\tau_1+\tau_2}$. The $U_{ij}$ represents the evolution under the $J_{ij}$ coupling interaction. We can now adjust the strength of $J_{ij}^{\text{eff}}$ by changing the intervals of $\tau_1$ and $\tau_2$. If $\tau_1 > \tau_2$, the $J_{ij}^{\text{eff}}$ term has the negative sign; if $\tau_1 = \tau_2$, the $J_{ij}^{\text{eff}}$ term is turned off, corresponding to the same sequence with the spin echo sandwich.

The $\sigma_i^x\sigma_{i+1}^x$ or $\sigma_i^y\sigma_{i+1}^y$ terms can be generated via rotating the $\sigma_i^z\sigma_{i+1}^z$ terms by a pair of $\pi/2$ hard pulses along the $y$ or $x$ axis. Therefore, combined with the above techniques, we realize the effective Hamiltonian of SSH model
\begin{equation}\label{eq: simHamofSSH}
	\mathcal{H}_{\text{SSH}}^{\text{S, eff}}=-\frac{\pi}{2}\sum_{i=1}^3J_{i,i+1}^{\text{eff}}(\sigma_i^x\sigma_{i+1}^x + \sigma_i^y\sigma_{i+1}^y).
\end{equation}
It can be set to the topological configuration with $J_{12}^{\text{eff}}, J_{34}^{\text{eff}}\approx 0 \text{Hz}, J_{23}^{\text{eff}}=69.7 \text{Hz}$, and the trivial configuration with $J_{12}^{\text{eff}}, J_{34}^{\text{eff}}=41.6 \text{Hz}, J_{23}^{\text{eff}}\approx 0 \text{Hz}$.

The external Hamiltonian of an introduced magnetic field $\mathcal{H}_{\text{ext}}$ can be realized by applying a hard RF pulse that acts on all carbon nuclei, whose corresponding Hamiltonian is 
\begin{equation}\label{eqs: drivenField}
	\mathcal{H}_{\text{RF}}=-\pi B_{\text{RF}} \left [\text{cos} \phi_{\text{RF}}\sum_{i=1}^4\sigma_i^x-\text{sin} \phi_{\text{RF}}\sum_{i=1}^4\sigma_i^y \right ],
\end{equation}
where $B_{\text{RF}}$ and $\phi_{\text{RF}}$ are the tunable amplitude and phase of RF pulse respectively. Under the rotation of a pair of $\pi/2$ hard pulses along $y$ axis, the $\sum_{i=1}^4\sigma_i^x$ in Eq. (\ref{eqs: drivenField}) becomes $\sum_{i=1}^4\sigma_i^z$. Finally, we obtain the target Hamiltonian 
\begin{equation}\label{eq: totSimHam}
	\mathcal{H}_T=-\pi B_{\text{RF}} \left [-\text{sin} \phi_{\text{RF}}\sum_{i=1}^4\sigma_i^y +\text{cos} \phi_{\text{RF}}\sum_{i=1}^4\sigma_i^z \right ] - \mathcal{H}_{\text{SSH}}^{\text{S, eff}},
\end{equation}
which can be mapped to the spin Hamiltonian (\ref{eqs: totHam}) of the SSH model in the external magnetic field by setting $h_r=\pi B_{\text{RF}}$, $\theta=2\pi-\phi_{\text{RF}}$, and $\phi=\pi/2$.

\section{Numerical simulation and analysis}

\subsection{Adiabatic preparation of many-body ground states} \label{subsec: adiabaticPreparation}

\begin{figure}
\centering
\includegraphics[width = 0.95 \linewidth]{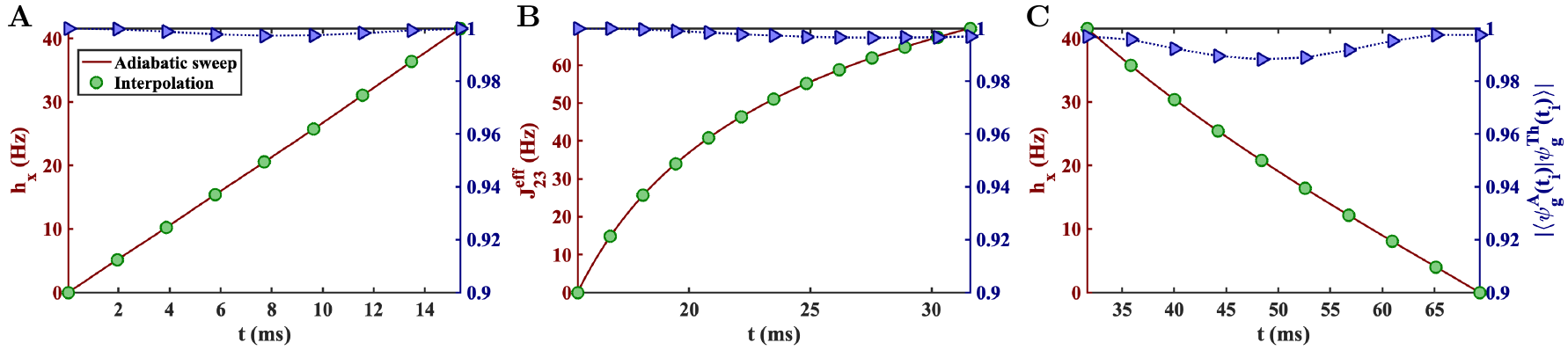}
\caption{ Adiabatic sweeps of $h_x$s and $J_{23}^{\text{eff}}$ along the Path 1 (shown in Fig. 1(D) of main text) for preparing the ground state,  corresponding to a topological phase.  The red solid lines were calculated according to the adiabatic condition (see Eq. (\ref{eqs: adiabaticCondition})). The green circles are the interpolations along the adiabatic sweeps.  The blue triangles represent the fidelities of $|\langle\psi_g^{\text{A}}(t_i)|\psi_g^{\text{Th}}(t_i)\rangle|$ between the theoretical and simulated ground states during the discretized time steps. When the evolution time T = 69.3 ms, the final fidelity is 0.998, which indicates the adiabatic preparation of ground state is receivable.}\label{figs: adiabaticSweep}
\end{figure}

The initial ground states with different configuration of SSH model were prepared adiabatically along the three paths shown in Fig. 1D of main text. Taking the Path 1 for example, the adiabatic sweeps of $h_x$s and $J_{23}^{\text{eff}}$ for preparing the topological phase are shown in Fig. \ref{figs: adiabaticSweep}, where the auxiliary sweeps of the external field along x direction $h_x$ are added to avoid the closed gap when scanning the strength of $J_{23}^{\text{eff}}$. We start with the initial state $|0000\rangle$, being the ground state of $\mathcal{H}_z = -h_z\sum_{i=1}^4\sigma_i^z$, and then sweeps the control parameters successively as follows: i) $h_x: 0 \rightarrow 41.6$Hz; ii) $J_{23}^{\text{eff}}: 0 \rightarrow 69.7$Hz; iii) $h_x: 41.6 \text{Hz} \rightarrow 0$. To ensure that the system always stays in the instantaneous ground state, the control parameters need to change slowly enough to satisfy the adiabatic condition \cite{Messiah1976Quantum}
\begin{equation}\label{eqs: adiabaticCondition}
	\left | \frac{\langle\psi_g|\dot{\psi}_e\rangle}{E_e - E_g} \right | = \frac{d |s(t)|}{dt}\left |\frac{\langle\psi_g|\partial_s \mathcal{H}(s)|\psi_e\rangle}{(E_e - E_g)^2}\right| \ll 1.
\end{equation}
Here the $|\psi_g\rangle$ and $|\psi_e\rangle$ represent the instantaneous ground and first excited states, and $s(t)$ denotes the shorthand symbol of control parameters of $h_x$ and $J_{23}^{\text{eff}}$. The above condition determines the optimal sweep of control parameters of $h_x$ and $J_{23}^{\text{eff}}$, shown in the solid red lines of Fig. \ref{figs: adiabaticSweep}. In experiments, the sweeps were discretized 32 steps (green circles), and the total time is $T = 69.3$ ms. The right longitudinal axis represents the fidelity between the theoretical and simulated ground states, that is, $|\langle\psi_g^{\text{A}}(t_i)|\psi_g^{\text{Th}}(t_i)\rangle|$. It can be found that the numerical value of fidelity for each time step is above 0.99, and final fidelity reaches to 0.998. That indicates it is credible that the system always stay in the ground state in the adiabatic evolution.

\subsection{Optimization of the quenching time}
 
\begin{figure}
\centering
\includegraphics[width = 0.95 \linewidth]{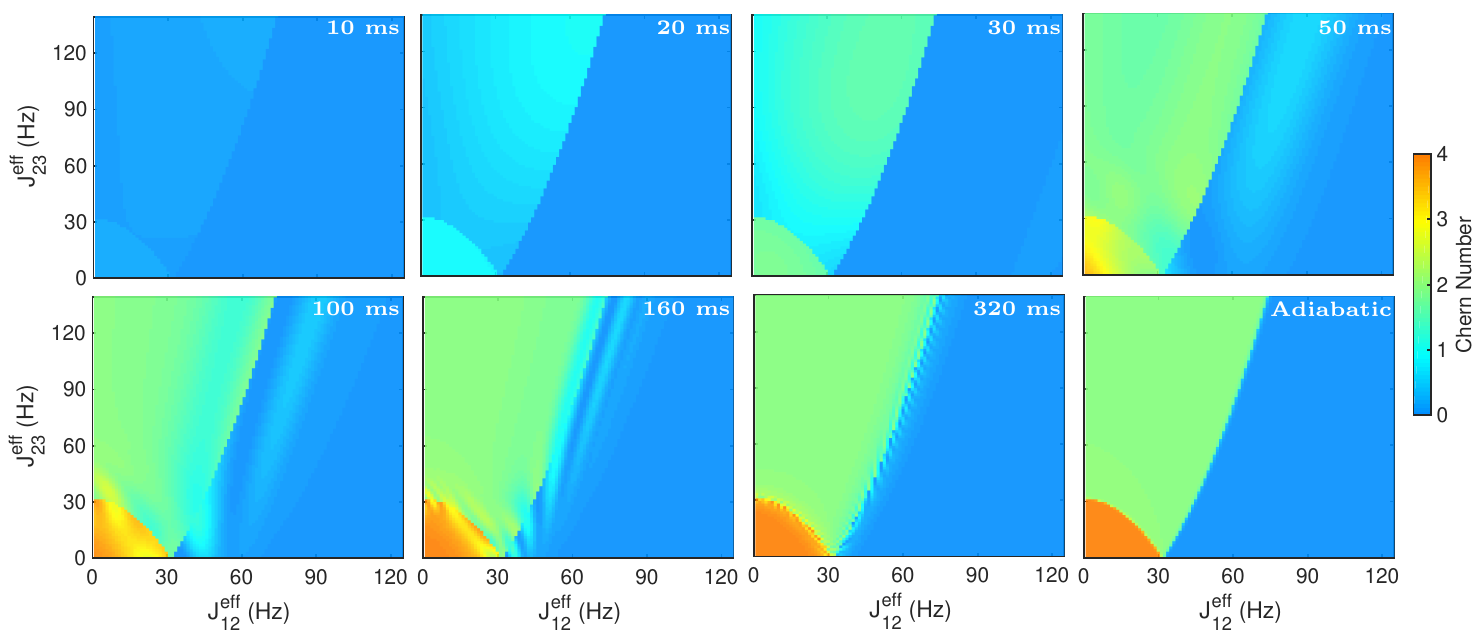}
\caption{Chern number numerically extracted from the linear response at different quench times $t_f$, where $t_f$ is the time it takes to ramp from the north pole of spherical manifold to its south pole. In these simulations, the external magnetic field was fixed to $h_r = 50$ Hz. The longer $t_f$ is more adiabatic and hence better in the simulation results. The final picture (labelled by ``Adiabatic'') is plotted by theoretical calculation. }\label{figs: optimizedTime}
\end{figure}

In order to extract the Berry curvatures with the high accuracy from the linear responses, the quench velocity $v_{\theta}$ should be slow enough such that the high-order terms of $v_{\theta}$ are negligible. We simulated the dynamical responses of the generalized force at different quench time $t_f = \pi/v_{\theta}$, with fixed $h_r = 50$ Hz. The $t_f$ is the time it takes to ramp from the north pole of spherical manifold of $\vec{h}$ to its south pole. The numerical Chern numbers are shown in Fig. \ref{figs: optimizedTime}, where the $t_f$ becomes longer, the dynamical quench is more adiabatic and the phase diagram approaches to the theoretical one (the final picture in Fig. \ref{figs: optimizedTime}) calculated by the definition of Chern number. When $t_f = 320$ ms, it is a good simulation with the theoretical calculation. Therefore, the quench time $t_f = 350$ ms taken in our experiments was indeed slow enough, such that the high-order errors remain small.

\subsection{Robustness analysis of degeneracy points}
 
\begin{figure}
\centering
\includegraphics[width = 1 \linewidth]{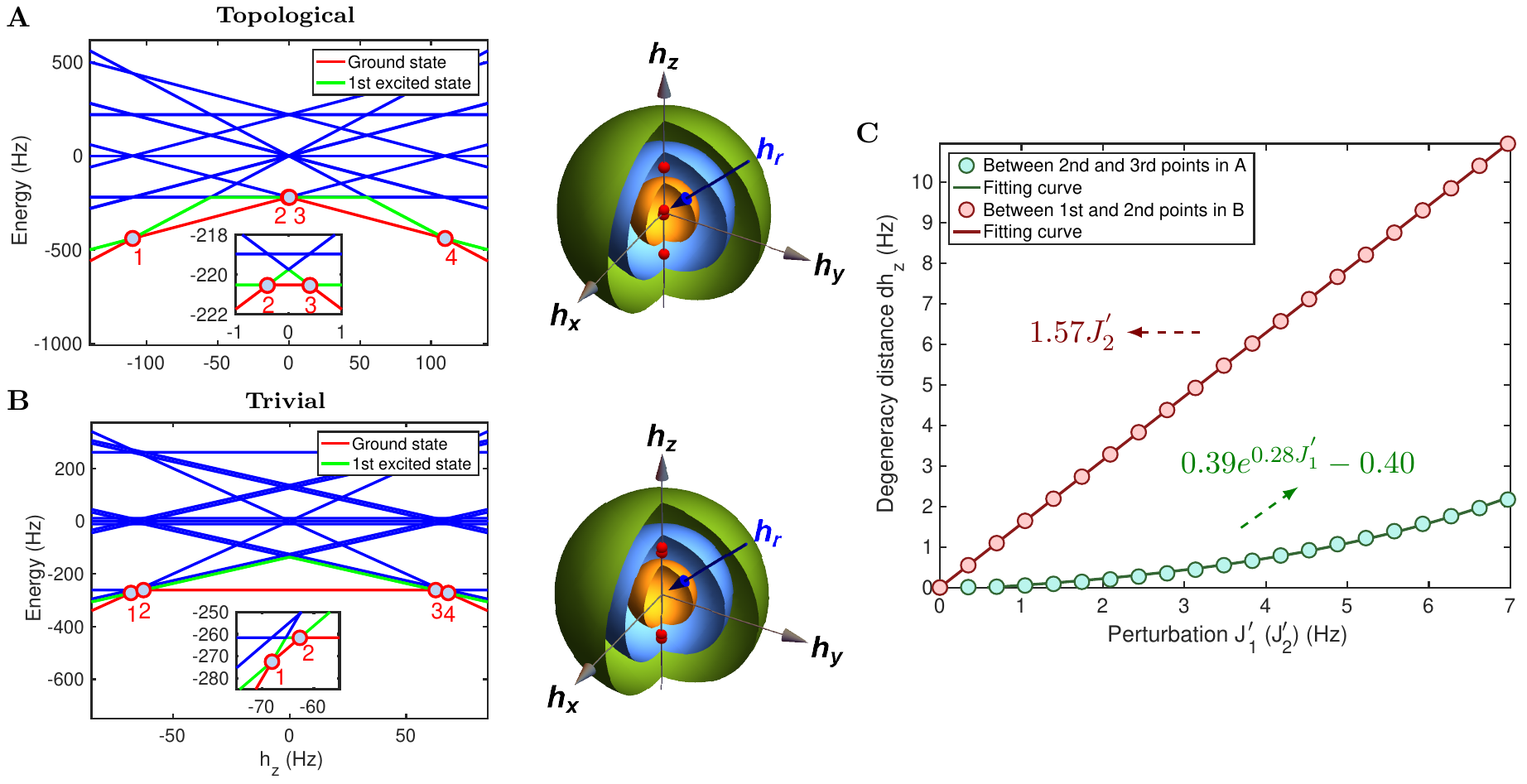}
\caption{Energy-level diagram and degeneracy points of topological (A) and trivial (B) configurations in the presence of small perturbations. The degeneracy points enclosed in the spherical field act as the sources of Chern number and are analogous to magnetic monopoles. (C) Robustness of degeneracy points against the perturbations of intracell hopping $J_1^{'}$ or intercell hopping $J_2^{'}$. The green (red) circles represent the degeneracy distance $dh_z$ between the second and the third (the first and the second) points in the topological (trivial) energy-level diagram. The fitting results show that topological degeneracies (separate exponentially) are more robust than trivial ones (separate linearly) under small perturbations. }\label{figs: degeneracyRobustness}
\end{figure}

In Subsection (\ref{subsec: ChernNumber}), we have explained the meaning of Chern number, which reveals the singularities of systems and corresponds to the degeneracy points enclosed in the spherical parameter manifold of $\vec{h}$. The nonzero value of Chern number emerges in the integral form of Eq. (\ref{eqs: BerryCurvature}) if and only if the ground states is degenerate. For the small systems with four qubits, the energy-level diagrams of the topological and trivial configurations in the presence of small perturbations are readily plotted by numerical calculations, and we can then obtain the locations of points of degenerate ground states, shown in Figs \ref{figs: degeneracyRobustness}A and \ref{figs: degeneracyRobustness}B. The small perturbations are introduced here for visualizing the number of degeneracy points more intuitively. We further analyze the robustness of topological and trivial phases by the numerical calculations of the distance $dh_z$ between the degeneracy points. In the presence of intracell hopping perturbation $J_1^{'}$, the topological degeneracies between the second and the third points in Fig. \ref{figs: degeneracyRobustness}A separate exponentially (see the green circles with the fitting curve of $dh_z = 0.39 e^{0.28 J_1^{'}} - 0.40$ in Fig. \ref{figs: degeneracyRobustness}C), whereas in the presence of intercell hopping perturbation $J_2^{'}$, the trivial degeneracies between the first and the second points in Fig. \ref{figs: degeneracyRobustness}B separate linearly (see the red circles with the fitting curve of $dh_z = 1.57 J_2^{'}$ in Fig. \ref{figs: degeneracyRobustness}C). The results show that topological degeneracies are more robust than trivial ones under small perturbations.

\section{Measurement and discussion}

\subsection{Measurement of spin-spin correlations}

\begin{figure}
\centering
\includegraphics[width = 0.72 \linewidth]{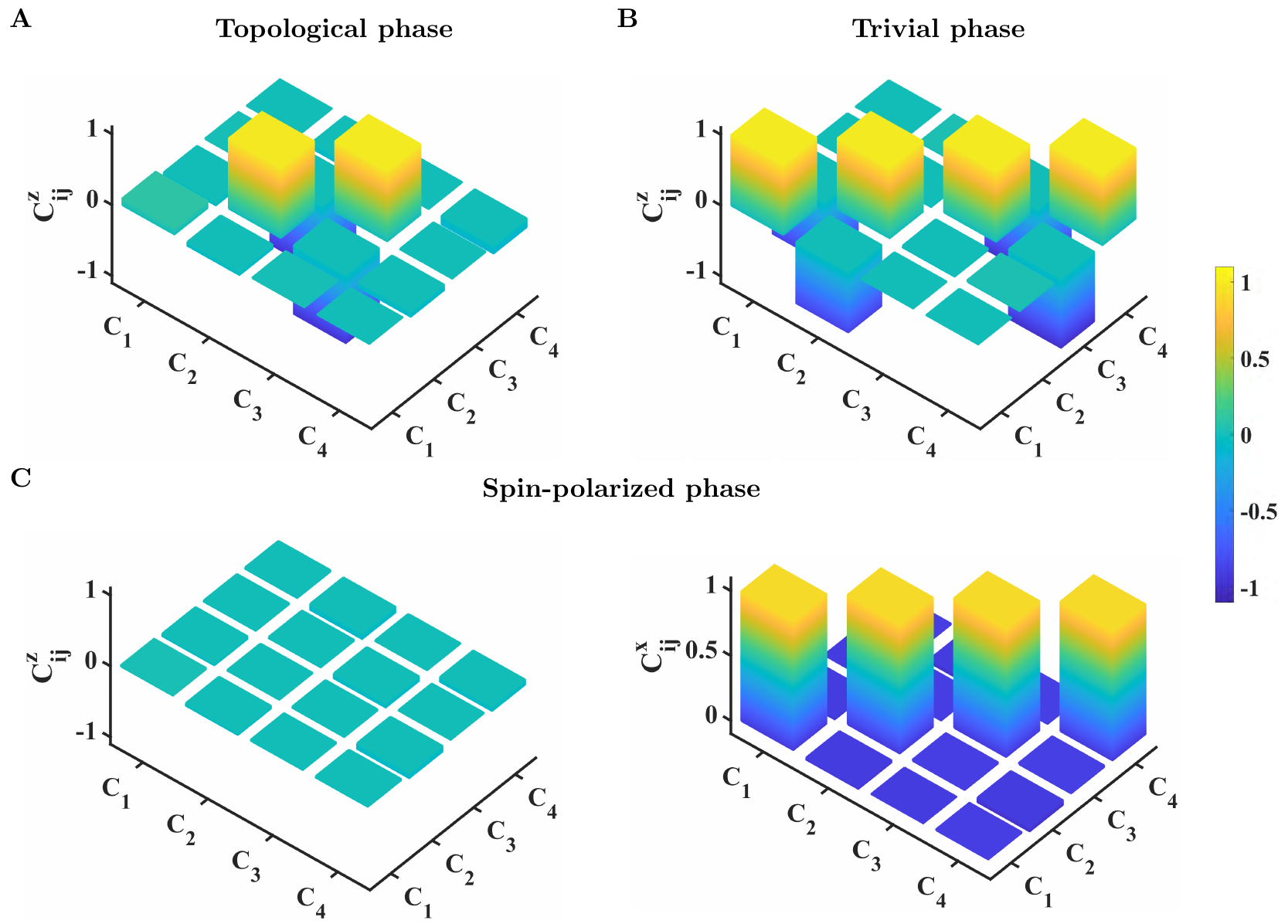}
\caption{Measured spin-spin correlations along the $z$ direction of (A) topological and (B) trivial phases. (C) show the measured spin-spin correlations along the $z$ (left) and $x$ (right) directions of spin-polarized phase. }\label{figs: correlation}
\end{figure}

We performed the measurements of the spin-spin correlations for different initial states using five independent readout experiments. The NMR signals obtained through the quadrature detection are given by \cite{Lee2002The},
\begin{equation}\label{eq:NMRsignal}
    S(t)=\text{Tr}\left[\rho_g R^{\dag}e^{i{H}_{\text{NMR}}t}\sum_{j=1}^4(\sigma_j^x+i\sigma_j^y)e^{-i{H}_{\text{NMR}}t}R\right],
\end{equation}
where $\rho_g$ represents the initial ground states prepared by adiabatic evolutions (see Subsection \ref{subsec: adiabaticPreparation}), and $R$ denotes the five readout pulses listed in Table \ref{tab:readout}, where $I$ is the unit operator. The corresponding readout spin operators can be used to reconstruct the spin-spin correlations of $C_{ij}^x = \langle\sigma_i^x\sigma_j^x\rangle-\langle\sigma_i^x\rangle\langle\sigma_j^x\rangle$ and $C_{ij}^z = \langle\sigma_i^z\sigma_j^z\rangle-\langle\sigma_i^z\rangle\langle\sigma_j^z\rangle$ for $i, j = 1, 2, 3, 4$. Figures \ref{figs: correlation}A and \ref{figs: correlation}B show the measured $C_{ij}^z$ of the topological and trivial phases. Combined with the results of $C_{ij}^x$ in Figs. 1E and 1F of main text, it further confirms the intercell spin correlations with $C_{23} = -0.969$ for the topological phase, and the intracell spin correlations with $C_{12}^z = -0.863, C_{34}^z = -0.984$ for the trivial phase. Different from the two topologically distinct phases, the measured $C_{ij}^z$ (Fig. \ref{figs: correlation}C, left) and $C_{ij}^x$ (Fig. \ref{figs: correlation}C, right) of spin-polarized (SP) phase demonstrate that there exist no correlation between spins. 

\begin{table}
\caption{Readout pulses and their corresponding readout spin operators for reconstructing the spin-spin correlations.}
\label{tab:readout}
  \centering
  \begin{tabular}{cc}
  \hline\hline
   Readout pulses & Readout spin operators \\
   \hline
   $I$ & $\sigma_1^x, \sigma_2^x, \sigma_3^x, \sigma_4^x$\\
   $R_1^y(\pi/2)$ & $\sigma_1^z, \sigma_1^z\sigma_2^z, \sigma_1^z\sigma_3^z, \sigma_1^z\sigma_4^z, \sigma_1^x\sigma_2^x, \sigma_1^x\sigma_3^x, \sigma_1^x\sigma_4^x$\\
   $R_2^y(\pi/2)$ & $\sigma_2^z, \sigma_2^z\sigma_3^z, \sigma_2^z\sigma_4^z, \sigma_2^x\sigma_3^x, \sigma_2^x\sigma_4^x$\\
   $R_3^y(\pi/2)$ & $\sigma_3^z, \sigma_3^z\sigma_4^z, \sigma_3^x\sigma_4^x$\\
   $R_4^y(\pi/2)$ & $\sigma_4^z$\\
   \hline\hline
  \end{tabular}
\end{table}

\subsection{Observation of the spin-polarized phase}

\begin{figure}
\centering
\includegraphics[width = 0.95 \linewidth]{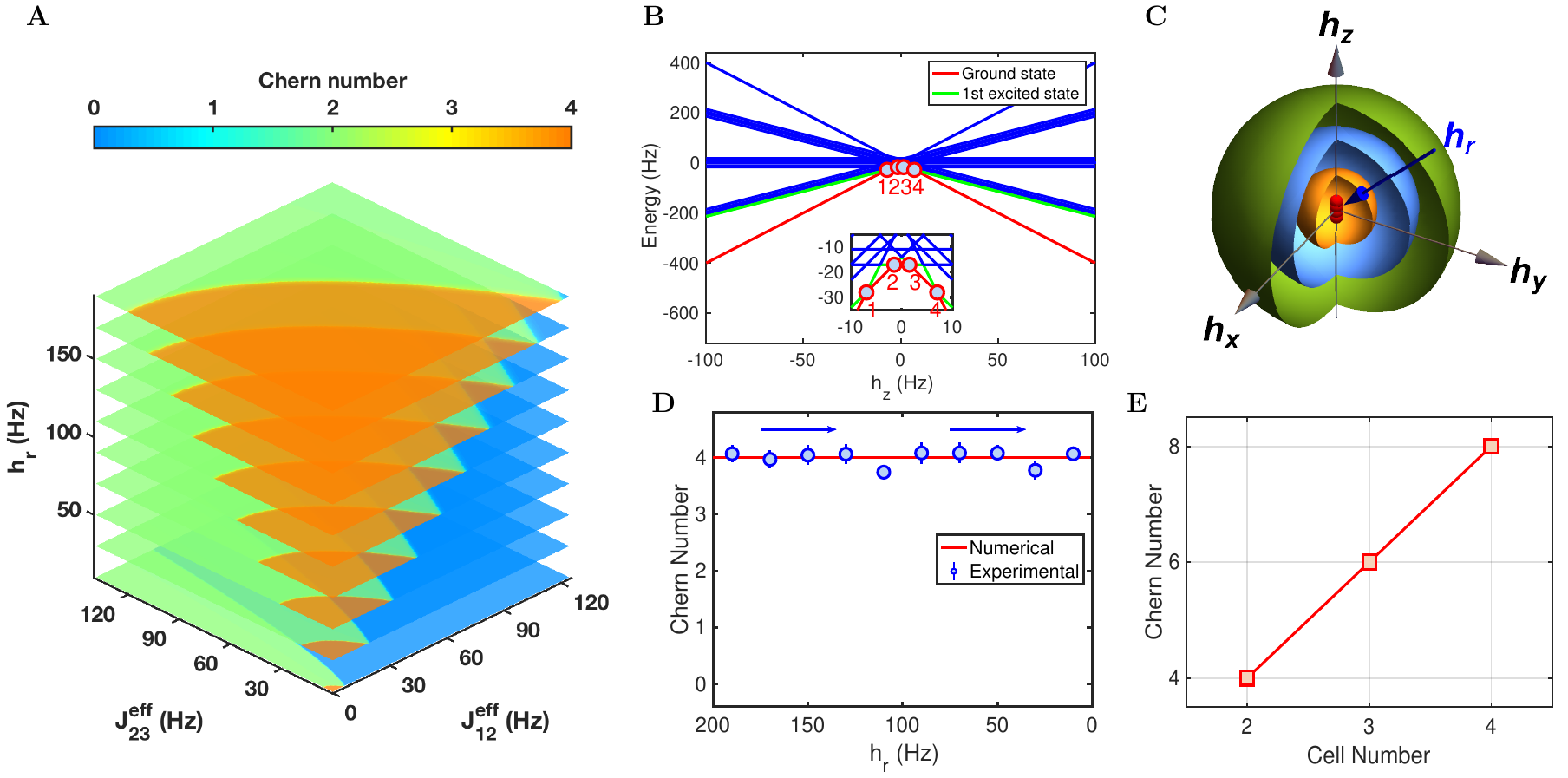}
\caption{(A) Calculated Chern number as a function of effective couplings $J_{12}^{\text{eff}}$ and $J_{23}^{\text{eff}}$ for different external magnetic fields $h_r$. The pink region of SP phase results from the external magnetic field and gradually decreases as $h_r \rightarrow 0$. (B) Energy-level diagram and degeneracy points of the SP configuration (i.e., $J_{12}^{\text{eff}}, J_{23}^{\text{eff}}, J_{34}^{\text{eff}} = 0$ Hz) in the presence of small perturbations. (C) Schematic of measuring the Chern number that corresponds to the total number of degeneracy points (small red balls) enclosed in spherical parameter space. The four degeneracy points at $h_r = 0$ totally originate from the introduced external magnetic field itself. (D) Measured Chern number in the SP configuration under different external magnetic fields, which was obtained from the integral of Berry curvature $\mathcal{F}_{\theta\phi}$ of Fig. \ref{figs: XjBC_fm}B. There is no obvious change of Chern number ($3.996\pm 0.065$), as $h_r \rightarrow 0$. (E) Calculated Chern number as a function of cell number. In the SP phases, the Chern number grows linearly with the increase of cell number.}\label{figs: ChernNumber_fm}
\end{figure}

We now discuss the influence of the introduced magnetic fields, which itself lead to the external region of SP phase in the absence of the intracell- and intercell hopping interactions of $J_{12}^{\text{eff}}$ and $J_{23}^{\text{eff}}$, as shown in Fig. \ref{figs: ChernNumber_fm}A. The pink region of SP phase depends on the field strength $h_r$, and gradually decreases to zero as $h_r \rightarrow 0$. The SP phase featuring the Chern number four can be understood from the energy-level diagram (Fig. \ref{figs: ChernNumber_fm}B). There are four points of degenerate ground states in the presence of small perturbations of $J_{12}^{\text{eff}}$ and $J_{23}^{\text{eff}}$, otherwise these points are superimposed without the perturbations. We know that there is a one-to-one correspondence between the number of degeneracy points and the Chern number (Fig. \ref{figs: ChernNumber_fm}C). When the system was set to the SP configuration (i.e., $J_{12}^{\text{eff}}, J_{23}^{\text{eff}}, J_{34}^{\text{eff}} = 0$ Hz), we measured the Chern number under different external magnetic fields shown in Fig. \ref{figs: ChernNumber_fm}D, and obtained the average Chern number $\mathcal{C}h = 3.996\pm 0.065$. There is no obvious change of Chern number as $h_r \rightarrow 0$, because the degeneracy points appear near $h_r = 0$. To gain further insight of the SP phase with the ``nontrivial'' Chern number, we analyze the scaling behavior. Figure \ref{figs: ChernNumber_fm}D shows the calculated Chern number dependence of system size, where the Chern number grows linearly with the increase of cell number for the SP phase which originates from the induced external magnetic fields.

\subsection{Other data for measuring the Chern numbers}

\begin{figure}
\centering
\includegraphics[width = 1 \linewidth]{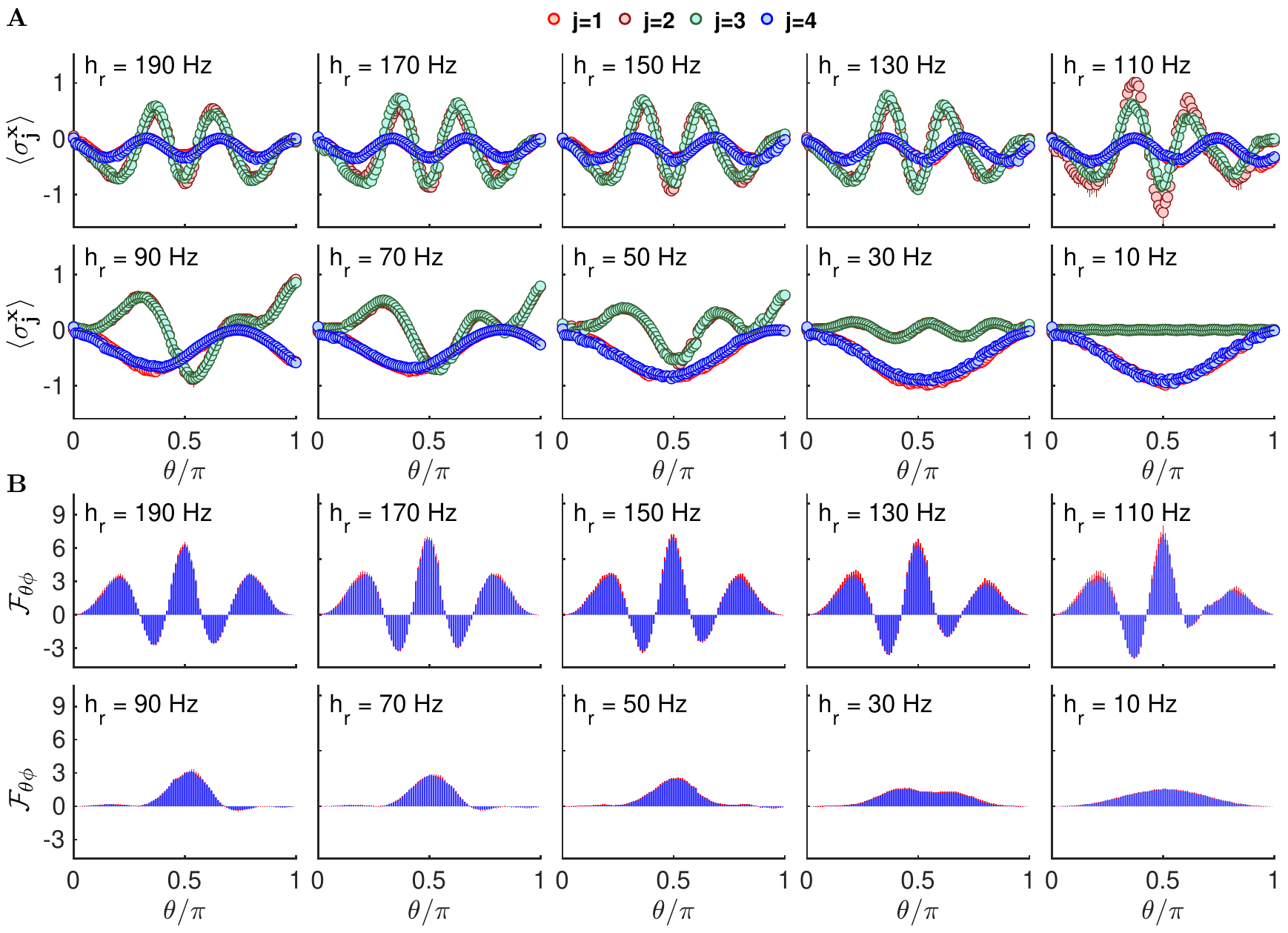}
\caption{(A) Measured spin magnetizations of $\langle\sigma_j^x\rangle$ for $j=1, 2, 3, 4$ under different external magnetic fields,  in the case of topological configuration, i.e., $J_{12}^{\text{eff}}, J_{34}^{\text{eff}} = 0$ Hz, $J_{23}^{\text{eff}} = 69.7$ Hz. Each $\langle\sigma_j^x\rangle$ was measured at 101 time steps during the quenching evolution. The solid lines show the numerical results.  (B) Extracted Berry curvatures from the linear responses of the spin magnetizations in (A). The error bars are given by the fitting uncertainties of experimental spectra, which refer to the readout errors.}\label{figs: XjBC_topo}
\end{figure}

\begin{figure}
\centering
\includegraphics[width = 1 \linewidth]{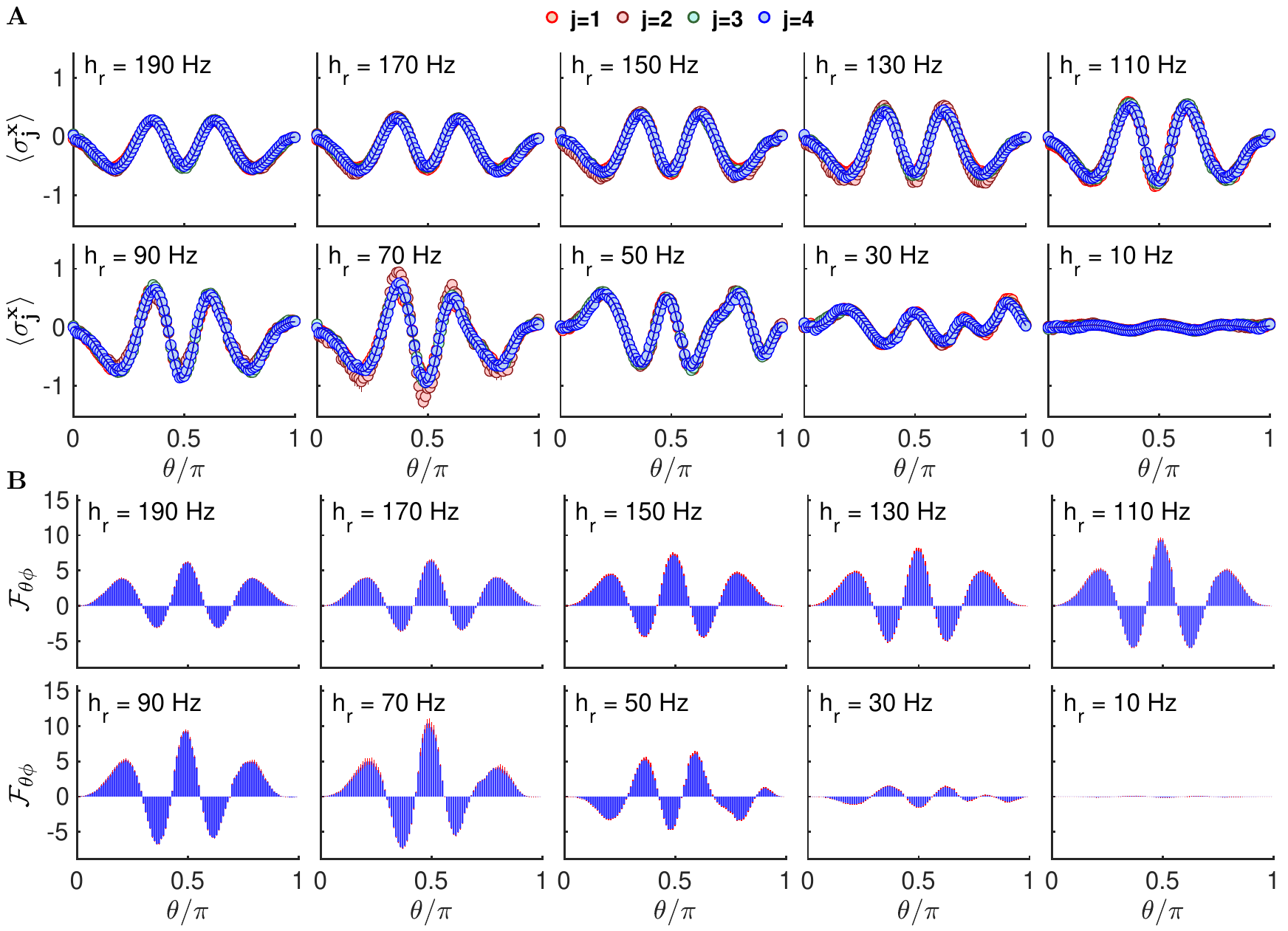}
\caption{(A) Measured spin magnetizations of $\langle\sigma_j^x\rangle$ for $j=1, 2, 3, 4$ under different external magnetic fields,  in the case of trivial configuration, i.e., $J_{12}^{\text{eff}}, J_{34}^{\text{eff}} = 41.6$ Hz, $J_{23}^{\text{eff}} = 0$ Hz. Each $\langle\sigma_j^x\rangle$ was measured at 101 time steps during the quenching evolution. The solid lines show the numerical results. (B) Extracted Berry curvatures from the linear responses of the spin magnetizations in (A). The error bars are given by the fitting uncertainties of experimental spectra, which refer to the readout errors.}\label{figs: XjBC_trivial}
\end{figure}

\begin{figure}
\centering
\includegraphics[width = 1 \linewidth]{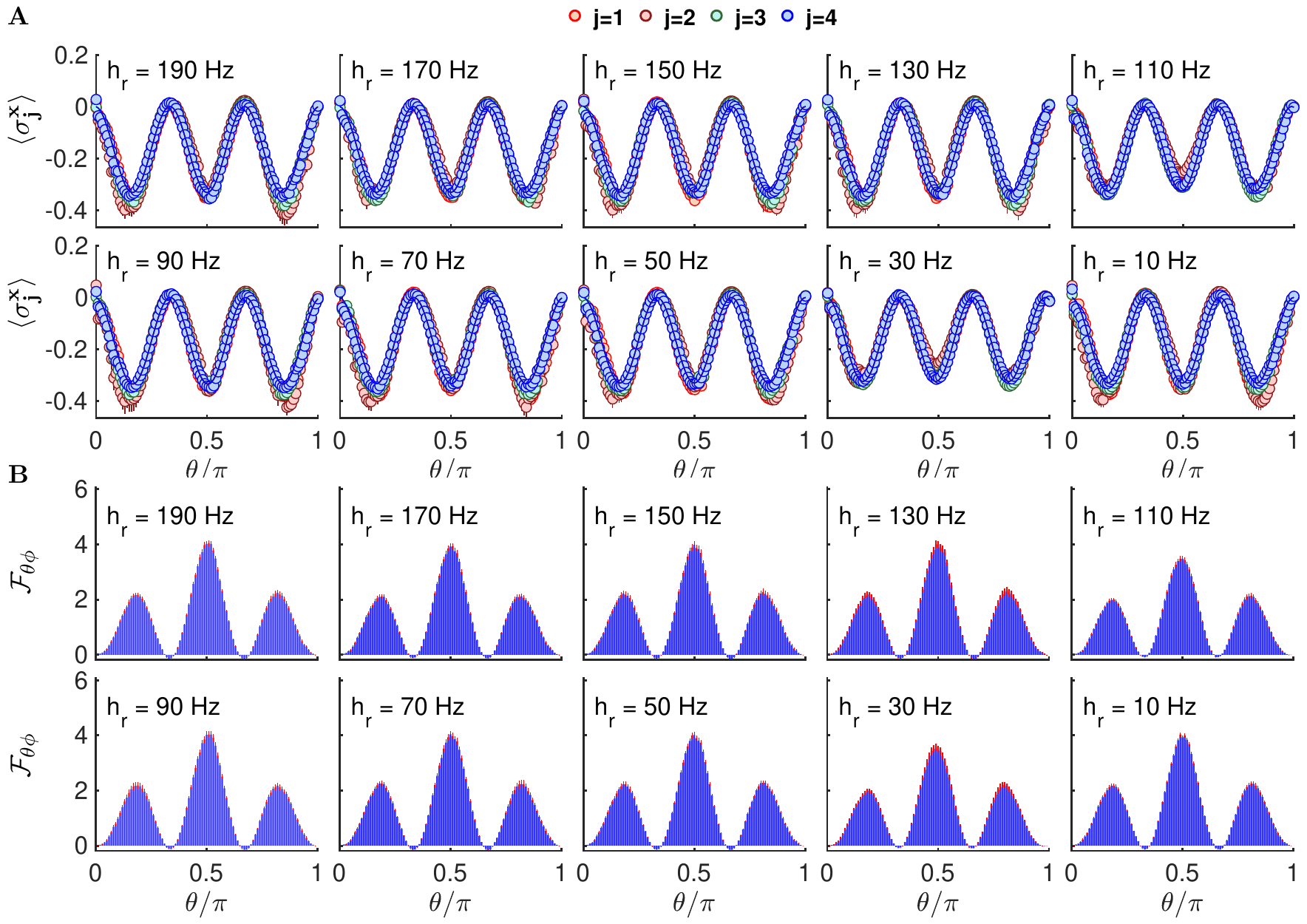}
\caption{(A) Measured spin magnetizations of $\langle\sigma_j^x\rangle$ for $j=1, 2, 3, 4$,  in the case of spin-polarized configuration, i.e., $J_{12}^{\text{eff}}, J_{34}^{\text{eff}}, J_{23}^{\text{eff}} = 0$ Hz. The initial states for different external magnetic fields $h_r$ began with $|0000\rangle$, corresponding to the ground states at the north pole of parameter spheres, and each $\langle\sigma_j^x\rangle$ was measured at 101 time steps during the quenching evolution. The solid lines show the numerical results. (B) Extracted Berry curvatures from the linear responses of the spin magnetizations in (A). The error bars are given by the fitting uncertainties of experimental spectra, which refer to the readout errors.}\label{figs: XjBC_fm}
\end{figure}

\begin{figure}
\centering
\includegraphics[width = 1 \linewidth]{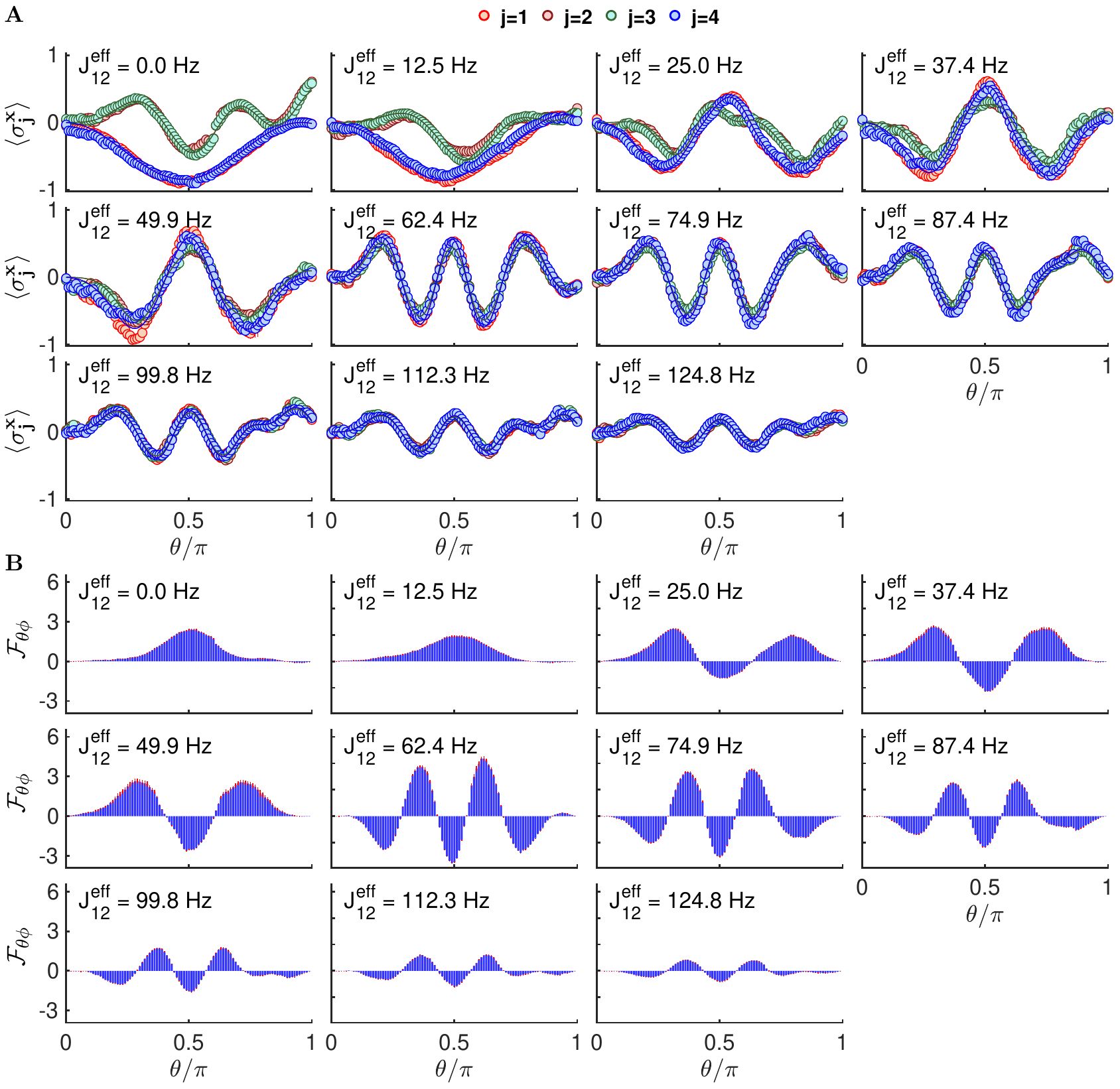}
\caption{(A) Measured spin magnetizations of $\langle\sigma_j^x\rangle$ for $j=1, 2, 3, 4$ under different effective couplings $J_{12}^{\text{eff}}$, with fixed $h_r = 50$ Hz and $J_{23}^{\text{eff}} = 69.7$ Hz. Each $\langle\sigma_j^x\rangle$ was measured at 101 time steps during the quenching evolution. The solid lines show the numerical results. (B) Extracted Berry curvatures from the linear responses of the spin magnetizations in (A). The error bars are given by the fitting uncertainties of experimental spectra, which refer to the readout errors.}\label{figs: XjBC_trans}
\end{figure}

The experimental Chern number of each $h_r$ was obtained by the integral of the Berry curvatures, which were extracted from the linear responses of the spin magnetizations. For the case of topological configuration,  the measured spin magnetizations $\langle\sigma_j^x\rangle$ for $j = 1, 2, 3, 4$ and the resulting Berry curvatures $\mathcal{F}_{\theta\phi}$ under different external magnetic fields are shown in Figs \ref{figs: XjBC_topo}A and \ref{figs: XjBC_topo}B, where each $\langle\sigma_j^x\rangle$ was measured 101 times along the quench path. The similar results for the cases of trivial and spin-polarized configurations are shown in Figs. \ref{figs: XjBC_trivial} and \ref{figs: XjBC_fm}, respectively. Figure \ref{figs: XjBC_trans} shows the experimental data of $\langle\sigma_j^x\rangle$ and $\mathcal{F}_{\theta\phi}$ under different effective couplings $J_{12}^{\text{eff}}$, with fixed $h_r = 50$ Hz and $J_{23}^{\text{eff}} = 69.7$ Hz, for probing the topological transition in the main text.

\subsection{Experimental error analysis}

We make an analysis to the data set of the case of topological configuration. The standard deviations of the experimental and simulated spin magnetizations were calculated via $\sigma_{\text{exp(sim)}} = \sqrt{\sum_{i=1}^M(\langle\sigma_j^x\rangle_{\text{exp(sim)}}-\langle\sigma_j^x\rangle_{\text{th}})^2/M}$, where $M=101$ represents the measured times. The experimental and simulated results for different external magnetic fields are listed in Table. \ref{tab: std}. It can be seen that the sources of experimental errors mainly include the imperfection of preparing the initial states (denoted as $\sigma_{\text{sim}}^{i}$), the inhomogeneity of quenching the RF fields (denoted as $\sigma_{\text{sim}}^{q}$), and the readout errors (denoted as $\sigma_{\text{sim}}^{\text{r}}$). Here the simulated standard deviations are investigated separately by only one error source, while the other procedures are assumed to be ideal. For the initial state preparation, the major errors are caused by the imperfection of the line-selective pulses in PPS $\rho_{0000}$ ($\approx 1.21\%$) and the adiabatic evolution of ground states at the north pole of spherical parameter manifold ($\approx 0.78\%$). The external magnetic field is realized by a hard RF pulse, which is inhomogeneous acting on all carbon nuclei with different amplitudes of $B_{\text{RF}}$ in Eq. (\ref{eqs: drivenField}). We simulated the quench dynamics based on a simple inhomogeneity model, which assume that the output amplitude discrepancy of RF field $B_{\text{RF}}$ is uniformly distributed between $\pm 3\%$. The average of simulated standard deviations of $\sigma_{\text{sim}}^{\text{q}}$ is approximately $1.50\%$. All measured values of $\langle\sigma_j^x\rangle$ in experiments are readout through the spectrum fitting. So the readout errors can be evaluated using the fitting uncertainties, which are determined by the absolute intensity of the spectral ground noise (white noise). It is worth noting that another source of error comes from the decoherence effect, however, which is small and can be ignored in our experiments, because the measured signals have been normalized using the initial ground states.
\begin{table}
\caption{Standard deviations of the experimental and simulated spin magnetizations under different external magnetic fields in the case of topological configuration. The $\sigma_{\text{sim}}^{\text{i}}, \sigma_{\text{sim}}^{\text{q}}$ and $\sigma_{\text{sim}}^{\text{r}}$ represent the standard deviations calculated by numerical simulations with only one error source, that is, the imperfection of preparing the initial states, the inhomogeneity of quenching the RF fields, and the readout errors, respectively.}
\label{tab: std}
  \centering
  \begin{tabular}{ccccccccccc}
  \hline\hline
   \ $h_r (\text{Hz})$ & 190 & 170 & 150 & 130 & 110 & 90 & 70& 50 & 30 & 10\\
   \hline
  \ $\sigma_{\text{exp}}$ & \ 3.83\%\  & \ 4.85\%\  & \ 4.22\%\  & \ 5.26\% & \ 8.17\% & \ 3.20\% & \ 3.73\% & \ 5.22\% & \ 3.46\% & \ 2.60\% \\
\  $\sigma_{\text{sim}}^{\text{i}}$ & \ 1.79\%\  & \ 2.10\%\  & \ 2.37\%\  & \ 2.41\% & \ 2.46\% & \ 1.96\% & \ 1.91\% & \ 1.78\% & \ 1.68\% & \ 1.59\% \\
\ $\sigma_{\text{sim}}^{\text{q}}$ & \ 1.72\%\  & \ 1.63\%\  & \ 2.19\%\  & \ 1.28\% & \ 1.08\% & \ 1.26\% & \ 1.52\% & \ 1.21\% & \ 1.76\% & \ 1.39\% \\
\ $\sigma_{\text{sim}}^{\text{r}}$ & \ 1.41\%\  & \ 1.64\%\  & \ 1.50\%\  & \ 2.00\% & \ 3.19\% & \ 1.04\% & \ 0.92\% & \ 1.18\% & \ 1.60\% & \ 1.19\% \\
   \hline\hline
  \end{tabular}
\end{table}


\begin{thebibliography}{10}
\expandafter\ifx\csname url\endcsname\relax
  \def\url#1{\texttt{#1}}\fi
\expandafter\ifx\csname urlprefix\endcsname\relax\def\urlprefix{URL }\fi
\providecommand{\bibinfo}[2]{#2}
\providecommand{\eprint}[2][]{\url{#2}}

\bibitem{Landau1937TheoryI}
\bibinfo{author}{Landau, L.~D.}
\newblock \bibinfo{title}{Theory of phase transformations}.
\newblock \emph{\bibinfo{journal}{Phys. Z. Sowjetunion}}
  \textbf{\bibinfo{volume}{11}}, \bibinfo{pages}{26} (\bibinfo{year}{1937}).

\bibitem{Wen1990Topological}
\bibinfo{author}{Wen, X.-G.}
\newblock \bibinfo{title}{Topological orders in rigid states}.
\newblock \emph{\bibinfo{journal}{International Journal of Modern Physics B}}
  \textbf{\bibinfo{volume}{4}}, \bibinfo{pages}{239--271}
  (\bibinfo{year}{1990}).

\bibitem{Luo2018Experimentally}
\bibinfo{author}{Luo, Z.} \emph{et~al.}
\newblock \bibinfo{title}{Experimentally probing topological order and its
  breakdown through modular matrices}.
\newblock \emph{\bibinfo{journal}{Nature Physics}}
  \textbf{\bibinfo{volume}{14}}, \bibinfo{pages}{160--165}
  (\bibinfo{year}{2018}).

\bibitem{Klitzing1980New}
\bibinfo{author}{Klitzing, K.~V.}, \bibinfo{author}{Dorda, G.} \&
  \bibinfo{author}{Pepper, M.}
\newblock \bibinfo{title}{New method for high-accuracy determination of the
  fine-structure constant based on quantized hall resistance}.
\newblock \emph{\bibinfo{journal}{Physical Review Letters}}
  \textbf{\bibinfo{volume}{45}}, \bibinfo{pages}{494--497}
  (\bibinfo{year}{1980}).

\bibitem{Hasan2010Colloquium}
\bibinfo{author}{Hasan, M.~Z.} \& \bibinfo{author}{Kane, C.~L.}
\newblock \bibinfo{title}{Colloquium: Topological insulators}.
\newblock \emph{\bibinfo{journal}{Reviews of Modern Physics}}
  \textbf{\bibinfo{volume}{82}}, \bibinfo{pages}{3045--3067}
  (\bibinfo{year}{2010}).

\bibitem{Qi2011Topological}
\bibinfo{author}{Qi, X.~L.} \& \bibinfo{author}{Zhang, S.~C.}
\newblock \bibinfo{title}{Topological insulators and superconductors}.
\newblock \emph{\bibinfo{journal}{Reviews of Modern Physics}}
  \textbf{\bibinfo{volume}{83}}, \bibinfo{pages}{1057--1110}
  (\bibinfo{year}{2011}).

\bibitem{Gu2009Tensor}
\bibinfo{author}{Gu, Z.~C.} \& \bibinfo{author}{Wen, X.~G.}
\newblock \bibinfo{title}{Tensor-entanglement-filtering renormalization
  approach and symmetry-protected topological order}.
\newblock \emph{\bibinfo{journal}{Physical Review B}}
  \textbf{\bibinfo{volume}{80}}, \bibinfo{pages}{155131}
  (\bibinfo{year}{2009}).

\bibitem{Chen2012Symmetry}
\bibinfo{author}{Chen, X.}, \bibinfo{author}{Gu, Z.~C.}, \bibinfo{author}{Liu,
  Z.~X.} \& \bibinfo{author}{Wen, X.~G.}
\newblock \bibinfo{title}{Symmetry-protected topological orders in interacting
  bosonic systems}.
\newblock \emph{\bibinfo{journal}{Science}} \textbf{\bibinfo{volume}{338}},
  \bibinfo{pages}{1604--1606} (\bibinfo{year}{2012}).

\bibitem{Wang2009Observation}
\bibinfo{author}{Wang, Z.}, \bibinfo{author}{Chong, Y.},
  \bibinfo{author}{Joannopoulos, J.~D.} \& \bibinfo{author}{Solja{\v{c}}i{\'c},
  M.}
\newblock \bibinfo{title}{Observation of unidirectional backscattering-immune
  topological electromagnetic states}.
\newblock \emph{\bibinfo{journal}{Nature}} \textbf{\bibinfo{volume}{461}},
  \bibinfo{pages}{772--775} (\bibinfo{year}{2009}).

\bibitem{Rechtsman2013Photonic}
\bibinfo{author}{Rechtsman, M.~C.} \emph{et~al.}
\newblock \bibinfo{title}{Photonic floquet topological insulators}.
\newblock \emph{\bibinfo{journal}{Nature}} \textbf{\bibinfo{volume}{496}},
  \bibinfo{pages}{196--200} (\bibinfo{year}{2013}).

\bibitem{Susstrunk2015Observation}
\bibinfo{author}{S{\"u}sstrunk, R.} \& \bibinfo{author}{Huber, S.~D.}
\newblock \bibinfo{title}{Observation of phononic helical edge states in a
  mechanical topological insulator}.
\newblock \emph{\bibinfo{journal}{Science}} \textbf{\bibinfo{volume}{349}},
  \bibinfo{pages}{47--50} (\bibinfo{year}{2015}).

\bibitem{Hafezi2013Imaging}
\bibinfo{author}{Hafezi, M.}, \bibinfo{author}{Mittal, S.},
  \bibinfo{author}{Fan, J.}, \bibinfo{author}{Migdall, A.} \&
  \bibinfo{author}{Taylor, J.~M.}
\newblock \bibinfo{title}{Imaging topological edge states in silicon
  photonics}.
\newblock \emph{\bibinfo{journal}{Nature Photonics}}
  \textbf{\bibinfo{volume}{7}}, \bibinfo{pages}{1001--1005}
  (\bibinfo{year}{2013}).

\bibitem{Meier2016Observation}
\bibinfo{author}{Meier, E.~J.}, \bibinfo{author}{An, F.~A.} \&
  \bibinfo{author}{Gadway, B.}
\newblock \bibinfo{title}{Observation of the topological soliton state in the
  su-schrieffer-heeger model}.
\newblock \emph{\bibinfo{journal}{Nature Communications}}
  \textbf{\bibinfo{volume}{7}} (\bibinfo{year}{2016}).

\bibitem{Leder2016Real-space}
\bibinfo{author}{Leder, M.} \emph{et~al.}
\newblock \bibinfo{title}{Real-space imaging of a topologically protected edge
  state with ultracold atoms in an amplitude-chirped optical lattice}.
\newblock \emph{\bibinfo{journal}{Nature Communications}}
  \textbf{\bibinfo{volume}{7}}, \bibinfo{pages}{13112} (\bibinfo{year}{2016}).

\bibitem{Zhang2019Experimental}
\bibinfo{author}{Zhang, Z.~F.} \emph{et~al.}
\newblock \bibinfo{title}{Experimental realization of multiple topological edge
  states in a 1d photonic lattice}.
\newblock \emph{\bibinfo{journal}{Laser $\&$ Photonics Reviews}}
  \textbf{\bibinfo{volume}{13}}, \bibinfo{pages}{1800202}
  (\bibinfo{year}{2019}).

\bibitem{de2019Observation}
\bibinfo{author}{de~L{\'e}s{\'e}leuc, S.} \emph{et~al.}
\newblock \bibinfo{title}{Observation of a symmetry-protected topological phase
  of interacting bosons with rydberg atoms}.
\newblock \emph{\bibinfo{journal}{Science}} \textbf{\bibinfo{volume}{365}},
  \bibinfo{pages}{775--780} (\bibinfo{year}{2019}).

\bibitem{Flaschner2016Experimental}
\bibinfo{author}{Flaschner, N.} \emph{et~al.}
\newblock \bibinfo{title}{Experimental reconstruction of the berry curvature in
  a floquet bloch band}.
\newblock \emph{\bibinfo{journal}{Science}} \textbf{\bibinfo{volume}{352}},
  \bibinfo{pages}{1091--1094} (\bibinfo{year}{2016}).

\bibitem{Aidelsburger2015Measuring}
\bibinfo{author}{Aidelsburger, M.} \emph{et~al.}
\newblock \bibinfo{title}{Measuring the chern number of hofstadter bands with
  ultracold bosonic atoms}.
\newblock \emph{\bibinfo{journal}{Nature Physics}}
  \textbf{\bibinfo{volume}{11}}, \bibinfo{pages}{162--166}
  (\bibinfo{year}{2015}).

\bibitem{Li2016Bloch}
\bibinfo{author}{Li, T.} \emph{et~al.}
\newblock \bibinfo{title}{Bloch state tomography using wilson lines}.
\newblock \emph{\bibinfo{journal}{Science}} \textbf{\bibinfo{volume}{352}},
  \bibinfo{pages}{1094--1097} (\bibinfo{year}{2016}).

\bibitem{Atala2013Direct}
\bibinfo{author}{Atala, M.} \emph{et~al.}
\newblock \bibinfo{title}{Direct measurement of the zak phase in topological
  bloch bands}.
\newblock \emph{\bibinfo{journal}{Nature Physics}}
  \textbf{\bibinfo{volume}{9}}, \bibinfo{pages}{795--800}
  (\bibinfo{year}{2013}).

\bibitem{Wang2019Direct}
\bibinfo{author}{Wang, Y.} \emph{et~al.}
\newblock \bibinfo{title}{Direct observation of topology from single-photon
  dynamics}.
\newblock \emph{\bibinfo{journal}{Physical Review Letters}}
  \textbf{\bibinfo{volume}{122}}, \bibinfo{pages}{193903}
  (\bibinfo{year}{2019}).

\bibitem{Su1979Solitons}
\bibinfo{author}{Su, W.~P.}, \bibinfo{author}{Schrieffer, J.} \&
  \bibinfo{author}{Heeger, A.~J.}
\newblock \bibinfo{title}{Solitons in polyacetylene}.
\newblock \emph{\bibinfo{journal}{Physical review letters}}
  \textbf{\bibinfo{volume}{42}}, \bibinfo{pages}{1698} (\bibinfo{year}{1979}).

\bibitem{Asboth2016A}
\bibinfo{author}{Asb{\'o}th, J.~K.}, \bibinfo{author}{Oroszl{\'a}ny, L.} \&
  \bibinfo{author}{P{\'a}lyi, A.}
\newblock \bibinfo{title}{A short course on topological insulators}.
\newblock \emph{\bibinfo{journal}{Lecture notes in physics}}
  \textbf{\bibinfo{volume}{919}}, \bibinfo{pages}{166} (\bibinfo{year}{2016}).

\bibitem{Choo2018Measurement}
\bibinfo{author}{Choo, K.}, \bibinfo{author}{von Keyserlingk, C.~W.},
  \bibinfo{author}{Regnault, N.} \& \bibinfo{author}{Neupert, T.}
\newblock \bibinfo{title}{Measurement of the entanglement spectrum of a
  symmetry-protected topological state using the ibm quantum computer}.
\newblock \emph{\bibinfo{journal}{Phys Rev Lett}}
  \textbf{\bibinfo{volume}{121}}, \bibinfo{pages}{086808}
  (\bibinfo{year}{2018}).

\bibitem{Gritsev2012Dynamical}
\bibinfo{author}{Gritsev, V.} \& \bibinfo{author}{Polkovnikov, A.}
\newblock \bibinfo{title}{Dynamical quantum hall effect in the parameter
  space}.
\newblock \emph{\bibinfo{journal}{Proceedings of the National Academy of
  Sciences}} \textbf{\bibinfo{volume}{109}}, \bibinfo{pages}{6457--6462}
  (\bibinfo{year}{2012}).

\bibitem{Roushan2014Observation}
\bibinfo{author}{Roushan, P.} \emph{et~al.}
\newblock \bibinfo{title}{Observation of topological transitions in interacting
  quantum circuits}.
\newblock \emph{\bibinfo{journal}{Nature}} \textbf{\bibinfo{volume}{515}},
  \bibinfo{pages}{241--244} (\bibinfo{year}{2014}).

\bibitem{Schroer2014Measuring}
\bibinfo{author}{Schroer, M.} \emph{et~al.}
\newblock \bibinfo{title}{Measuring a topological transition in an artificial
  spin-1/2 system}.
\newblock \emph{\bibinfo{journal}{Physical review letters}}
  \textbf{\bibinfo{volume}{113}}, \bibinfo{pages}{050402}
  (\bibinfo{year}{2014}).

\bibitem{Luo2016Experimental}
\bibinfo{author}{Luo, Z.} \emph{et~al.}
\newblock \bibinfo{title}{Experimental observation of topological transitions
  in interacting multispin systems}.
\newblock \emph{\bibinfo{journal}{Physical Review A}}
  \textbf{\bibinfo{volume}{93}}, \bibinfo{pages}{052116}
  (\bibinfo{year}{2016}).

\bibitem{SM2}
\bibinfo{note}{Full details on the experimental and theoretical procedures are
  available as supplementary materials}.

\bibitem{Shaka1983Improved}
\bibinfo{author}{Shaka, A.}, \bibinfo{author}{Keeler, J.},
  \bibinfo{author}{Frenkiel, T.} \& \bibinfo{author}{Freeman, R.}
\newblock \bibinfo{title}{An improved sequence for broadband decoupling:
  Waltz-16}.
\newblock \emph{\bibinfo{journal}{Journal of Magnetic Resonance (1969)}}
  \textbf{\bibinfo{volume}{52}}, \bibinfo{pages}{335--338}
  (\bibinfo{year}{1983}).

\bibitem{Messiah1976Quantum}
\bibinfo{author}{Messiah, A.}
\newblock \bibinfo{note}{Quantum Mechanics (Wiley, New York, 1976)}.

\bibitem{Lee2002The}
\bibinfo{author}{Lee, J.-S.}
\newblock \bibinfo{title}{The quantum state tomography on an nmr system}.
\newblock \emph{\bibinfo{journal}{Physics Letters A}}
  \textbf{\bibinfo{volume}{305}}, \bibinfo{pages}{349--353}
  (\bibinfo{year}{2002}).

\end{thebibliography}
\end{document}